\documentclass[a4paper,11pt]{article}
\pdfoutput=1 

\usepackage{jcappub} 
\usepackage{mathtools}

\usepackage{array}
\newcolumntype{P}[1]{>{\centering\arraybackslash}p{#1}}
\newcolumntype{M}[1]{>{\centering\arraybackslash}m{#1}}
\usepackage{graphicx}
\usepackage[dvipsnames]{xcolor}
\usepackage{longtable}
\usepackage{url}
\usepackage{verbatim}
\usepackage[utf8]{inputenc}
\usepackage{makecell}
\usepackage{comment}
\usepackage{cleveref}
\crefname{section}{§}{§§}
\usepackage[T1]{fontenc} 
\graphicspath{ {Figures/} }

\title{\boldmath Spectral and spatial analysis of the dark matter subhalo candidates among \textit{Fermi} Large Area Telescope unidentified sources}

\author[a,b]{Javier Coronado-Bl\'azquez}
\author[a,b]{Miguel A. S\'anchez-Conde}
\author[c,d]{Mattia Di Mauro}
\author[a,b]{Alejandra Aguirre-Santaella}
\author[e]{Ioana Ciuc\u{a}}
\author[f]{Alberto Dom\'inguez}
\author[e]{Daisuke Kawata}
\author[c,g]{N\'estor Mirabal}

\affiliation[a]{Instituto de Física Teórica UAM-CSIC,\\Universidad Autónoma de Madrid, C/ Nicolás Cabrera, 13-15, 28049 Madrid, Spain}
\affiliation[b]{Departamento de Física Teórica, M-15,\\Universidad Autónoma de Madrid, E-28049 Madrid, Spain}
\affiliation[c]{NASA Goddard Space Flight Center, Greenbelt, MD 20771, USA}
\affiliation[d]{CRESST, Catholic University of America, Department of Physics, Washington DC 20064, USA}
\affiliation[e]{Mullard Space Science Laboratory, University College London, Holmbury St. Mary, Dorking, Surrey, RH5 6NT, UK}
\affiliation[f]{IPARCOS and Department of EMFTEL, Universidad Complutense de Madrid, E-28040 Madrid, Spain}
\affiliation[g]{CRESST/CSST/Department of Physics, UMBC, Baltimore, MD 21250, USA}

\emailAdd{javier.coronado@uam.es}
\emailAdd{miguel.sanchezconde@uam.es}
\newcommand{\Gaia}{\textit{Gaia}}

\abstract{\textit{Fermi}-LAT unidentified sources (unIDs) have proven to be compelling targets for performing indirect dark matter (DM) searches. In a previous work, we found that among the 1235 unIDs in \textit{Fermi}-LAT’s catalogs (3FGL, 2FHL and 3FHL) only 44 of those are DM subhalos candidates. We now implement a spectral analysis to test whether these remaining sources are compatible or not with DM origin. This analysis is executed using almost 10 years of Pass 8 \textit{Fermi}-LAT data. None of the unIDs are found to significantly prefer DM-induced emission compared to other, more conventional, astrophysical sources. In order to discriminate between pulsar and DM sources, we developed a new method which is based on the source’s spectral curvature, peak energy, and its detection significance. We also look for spatial extension, which may be a hint for a DM origin according to our N-body simulation studies of the subhalo population. In addition, we used $Gaia$ DR2 data to search for a potential stellar counterpart to our best DM subhalo candidates and, although no firm associations could be found, one of them coincides with the Sagittarius stream. Finally, previous constraints on the DM annihilation cross section are updated with the new number of remaining DM subhalo candidates among unIDs. Our limits now rule out canonical thermal WIMPs up to masses of 10 GeV for $b\bar{b}$ and 20 GeV for $\tau^+\tau^-$ annihilation channels, in this way being as sensitive and complementary to those obtained from other targets and probes.}

\begin{document}
\maketitle
\flushbottom

\section{Introduction}
\label{sec:intro}

During the last years, different search strategies have been employed in the quest for dark matter (DM) detection \cite{GarrettDuda09,Roszkowski+17}. Among these, we find direct detection experiments, designed to observe the elastic scattering of DM-nuclei interactions, DM production in accelerators such as the LHC, and indirect searches that look for the annihilation or decay products of DM. These approaches are in many cases complementary, allowing us to probe different classes of DM candidates, subject to different uncertainties.

Gamma rays have been hailed as the "golden channel" for indirect DM detection, as they are not deflected by magnetic fields as opposed to antimatter produced in DM annihilation or decays, and are easier to detect than neutrinos. With gamma-ray telescopes, many different observational targets have been pursued. Among the most promising sources are the Galactic center (GC) \cite{fermi_gc_paper16,fermi_gc_paper17, Abdallah2016, Pierre2014}, dwarf spheroidal galaxies (dSphs) \cite{dsphs_paper, Oakes2019, Ahnen16} and galaxy clusters \cite{fermi_cluster_paper}. Although the GC is expected to be the brightest source of gamma rays produced by DM annihilation, the astrophysical gamma-ray background, in the form of unresolved diffuse Galactic emission, is significant and poorly understood \cite{fermi_gc_pulsar_paper,Bartels2017}. In contrast, while the astrophysical background for dSphs is expected to be quite low, its DM annihilation signal is predicted to be comparatively fainter, and subject to large uncertainties due to the imprecise DM content modelling.

Another interesting and complementary target in gamma-ray searches are the dark satellites. Within the standard $\Lambda$CDM paradigm, the DM halos that host galaxies are expected to contain a large amount of substructure (subhalos) as a consequence of the bottom-up hierarchical formation history \cite{Madau2008}. The largest members of this subhalo population (above $\sim10^{7}$ M\textsubscript{\(\odot\)}) will host dSphs, yet most of them are not massive enough to retain baryonic (i.e. gas) content and form stars, and therefore should remain completely dark \cite{Walker13}. These objects are predicted to be compact and concentrated, and therefore, if near enough and composed of \textit{Weakly Interacting Massive Particles} (WIMPs), they would yield a significant flux of gamma rays that could be detectable by gamma-ray observatories such as \textit{Fermi}-LAT \cite{fermi_instrument_paper}. The main advantage of these sub-halos are that they are not expected to host any non-DM astrophysical gamma-ray emitters, their large number density in the Galaxy and their high concentrations of DM. The main disadvantages come from uncertainties in their structural properties, their exact number and their distribution in the Galaxy. These also translate into a total lack of knowledge on their sky position. Although subhalos are expected to be almost isotropically distributed from the Earth point of view \cite{Springel2008_nature}, in general terms, only those located close to us may yield annihilation fluxes large enough to be detected.

A large number of \textit{Fermi}-LAT detected sources lack proper astrophysical association and remain as unidentified sources (unIDs), therefore being perfect candidates to perform this search of Galactic DM subhalos or dark satellites. Many works have already used these targets for indirect DM detection \cite{Bertoni+15, Bertoni+16, Calore+17, Schoonenberg+16,HooperWitte17,BerlinHooper14, Zechlin+12, ZechlinHorns12, Belikov2012, BuckleyHooper10, fermi_dm_satellites_paper}. In a previous work \cite{Coronado_Blazquez2019} we performed a filtering of those unIDs present in the latest point-source catalogs (namely 3FGL \cite{3FGL_paper}, 2FHL \cite{2FHL_paper}, and 3FHL \cite{3FHL_paper}) according to the expected emission from DM subhalo annihilation. Then, we combined these catalog filtering results with a state-of-the-art N-body simulation, conveniently repopulated with low-mass subhalos below the original mass resolution limit, to set constraints on the DM annihilation cross section. To do so, it was necessary to perform a full characterization of the \textit{Fermi}-LAT sensitivity to DM annihilation for different annihilation channels and sky positions. The obtained constraints were competitive, ruling out canonical thermal WIMPs up to $\sim$6 GeV in the case of $\tau^+\tau^-$ annihilation.

In \cite{Coronado_Blazquez2019}, from a total of 1235 (1010 3FGL + 48 2FHL + 177 3FHL) unIDs, we were able to discard all but 44 unIDs (16 3FGL + 1 2FHL + 24 3FHL) as DM subhalos outside the Galactic plane ($|b|\geq10^\circ$). Although no spectral information was explicitly used, the filter based on machine learning results \cite{Lefaucheur2017,3fglzoo_paper} to reject unIDs as DM subhalo candidates implicitly did. However, we note that such filter referred to just a general and coarse analysis which searched for active galactic nuclei (AGN) features, such as the spectral index of the source, and only relying on the official, 4-year LAT data \cite{3FGL_paper}. In this work, we perform a full, dedicated spectral analysis of these 44 remaining candidates using all 10-year \textit{Fermi}-LAT available data, in order to confirm or discard their DM origin. With all this new spectral information at hand, we can reduce the number of potential subhalos among unIDs even further, to point out the best DM candidates, and to update our previous DM constraints, following the procedure described in \cite{Coronado_Blazquez2019} (see also \cite{Calore+17}). As it was shown in that analysis, the fewer the sources compatible with DM, the stronger the DM constraints.

In the process, it will be also necessary to address the long-standing problem of confusion between gamma-ray emission from pulsars and DM \cite{Mirabal2013,Mirabal2016}. This confusion is due to annihilating WIMP DM having similar spectra to most pulsars \cite{fermi_dm_satellites_paper} for low, $\mathcal{O}(10)$ GeV WIMP masses, especially for annihilation into $b\overline{b}$ quarks. As we will show in this work, a more clear separation between both classes of sources may be possible should the work be done in an optimal parameter space entirely built upon the available spectral information.

Additionally, spatial source extension has been pointed out to be a ``smoking gun'' for this kind of indirect DM searches \cite{Bertoni+16}. Therefore, we also perform a spatial analysis for the candidates surviving the DM spectral analysis. Should any of the considered sources exhibit significant extension, a DM origin would become a plausible option.

The structure of this paper is as follows: in \cref{sec:spectral_section}, we perform the spectral analysis and apply the relevant statistical tests needed to confidently evaluate the DM origin hypothesis. In \cref{sec:constraints}, we set DM constraints taking into account what is learnt from our spectral analysis. \cref{sec:beta} presents a promising new tool to discriminate between DM annihilation and pulsars. Our search for spatial extension on the most interesting candidates that we found after our spectral analysis is presented in \cref{sec:spatial_analysis}. A search for potential stellar streams in $Gaia$ DR2 within our best candidates is done in \cref{sec:Gaia_analysis}. We conclude our report in \cref{sec:conclusions}.

\section{Spectral analysis of DM candidates}
\label{sec:spectral_section}

\subsection{Technical setup and analysis pipeline}
\label{sec:tech_spectral}

The spectral analysis is performed with \textit{Fermipy} \cite{Fermipy_paper}. We use almost 10 years of data (MET\footnote{MET: \textit{Mission Ellapsed Time}} 239557417 to 541779795). The energy range is chosen to avoid photons under 300 MeV due to the intense diffuse emission, which would make the analysis very time-consuming. We use Pass 8 events \cite{pass8_paper}, with the {\tt P8R3\_SOURCE\_V2} instrumental response functions (IRFs, \cite{p8r3_paper}), and utilize all available photons (FRONT+BACK), excluding those arriving with zenith angles greater than $100^\circ$ below 1 GeV and $105^\circ$ above.

To compute the detection significance of the source, we adopt the Test Statistic, 
\begin{equation}
\label{eq:TS}
\mathrm{TS}=-2\cdot\textrm{log}\left[\frac{\mathcal{L}(H_1)}{\mathcal{L}(H_0)}\right]
\end{equation}

\noindent where $\mathcal{L}(H_0)$ and $\mathcal{L}(H_1)$ are respectively the likelihoods under the null (no source) and alternative (existing source) hypotheses. 

A region of interest (ROI) of $18^\circ\times18^\circ$ around the reported position is chosen for each source, with an angular bin size of $0.06^\circ$ and 8 (4) evenly spaced logarithmic energy bins for sources with TS>100 (TS<100). To model point sources lying within the ROI, we use the preliminary 8-year point source catalog, the so-called FL8Y\footnote{\url{https://fermi.gsfc.nasa.gov/ssc/data/access/lat/fl8y/}}. In this analysis, the {\tt gll\_iem\_v06.fits} diffuse and {\tt iso\_P8R3\_SOURCE\_V2.txt} isotropic templates are used\footnote{\url{https://fermi.gsfc.nasa.gov/ssc/data/access/lat/BackgroundModels.html}}. A summary of the analysis setup can be found in Table \ref{tab:analysis_setup}.

\begin{table}
  \begin{center}
    \begin{tabular}{|M{5cm}|M{6cm}|}
    \hline 
    Time domain (ISO 8601) & 2008-08-04 to 2018-03-03\\
    \hline 
    Time domain (MET) & 239557417 to 541779795\\
    \hline 
    Energy range & 300 MeV - 800 GeV\\
    \hline 
    IRF & {\tt P8R3\_SOURCE\_V2}\\
    \hline
    Event type & FRONT+BACK\\
    \hline
    Point-source catalog & FL8Y\\
    \hline
    ROI size & $18^\circ\times18^\circ$\\
    \hline
    Angular bin size & $0.06^\circ$\\
    \hline
    Bins per energy decade &  4 (TS<100) or 8 (TS>100)\\
    \hline
    Galactic diffuse model & {\tt gll\_iem\_v06.fits}\\
    \hline
    Isotropic diffuse model & {\tt iso\_P8R3\_SOURCE\_V2.txt}\\
    \hline
    \end{tabular}
    \caption{Summary of the spectral analysis technical setup. See \cref{sec:spectral_section} for details. }
    \label{tab:analysis_setup}
  \end{center}
\end{table}

First of all, we run {\tt GTAnalysis}\footnote{\url{https://fermipy.readthedocs.io/en/latest/}} and define the full analysis setup. Then, we perform the initial fit running the {\tt GTAnalysis.free\_sources}, {\tt GTAnalysis.fit} and {\tt GTAnalysis.optimize} modules. We remove sources under the $TS=25$ ($\sim5\sigma$) detection threshold and find new sources over it with {\tt GTAnalysis.find\_sources}. If the unID source is found to have $TS\geq25$, we recalculate its sky position with the {\tt GTAnalysis.localize} module. We also obtain a TS detection map of the whole ROI, removing the candidate source to verify that the residuals look good in the rest of the region and the found excess is indeed due to the unID. Also, the gamma-ray spectral energy distribution (SED) of the source is generated with the {\tt GTAnalysis.sed} module.

Now we perform the spectral fit using the three different parametric functions given below,

\begin{itemize}
\item
Power law (PL):

\begin{equation}
\label{eq:powerlaw}
\frac{dN}{dE}=K\left(\frac{E}{E_{0}}\right)^{-\Gamma}
\end{equation}

\noindent where $K$ is the normalization, $\Gamma$ is the photon spectral index and $E_0$ sets the energy scale. This form is widely used to fit gamma-ray AGNs and other sources.

\item
Log-parabola (LP):

\begin{equation}
\label{eq:logparabola}
\frac{dN}{dE}=K\left(\frac{E}{E_{0}}\right)^{-\Gamma-\beta\cdot \textrm{log}\left(E/E_0\right)}
\end{equation}

\noindent where, in addition to those parameters in Eq. \ref{eq:powerlaw}, the parameter $\beta$ or spectral curvature index is included. This parametric form is used to fit both blazars and pulsars, although it can also provide a good spectral description for DM in certain cases. Further details on $\beta$ in the context of DM searches will be discussed in \cref{sec:beta}.

\item
Power law with super-exponential cutoff (PLE):

\begin{equation}\label{eq:powerlaw_exponentialcutoff}
\frac{dN}{dE}=K\left(\frac{E}{E_{0}}\right)^{-\Gamma}e^{-\left(\frac{E}{E_{cut}}\right)^\beta}
\end{equation}

\noindent that includes $E_{cut}$ as the cutoff energy and $\beta$ as the super-exponential index. This spectral shape is normally used to fit pulsars, although it also fits well some possible DM annihilation spectra \cite{Coronado_Blazquez2019} \cite{Calore+17}. This fact reflects the above-mentioned spectral confusion between pulsars and DM \cite{fermi_dm_satellites_paper,Mirabal2013,Mirabal2016}.

\end{itemize}

The fit is performed with {\tt GTAnalysis.optimize} and {\tt GTAnalysis.fit} for these three parametric forms, leaving free the normalization of the sources within a radius of $2^\circ$ from the center of the ROI. We obtain the log-likelihood and the spectral parameters with their $1\sigma$ uncertainties. By comparing the log-likelihoods obtained for each of the three parametric forms, we are able to determine the best-fit model. In many cases, though, we anticipated that two or even the three of these models will yield similar likelihoods due to typically poor photon statistics.

\begin{figure}[!ht]
\centering
\includegraphics[height=4.4cm]{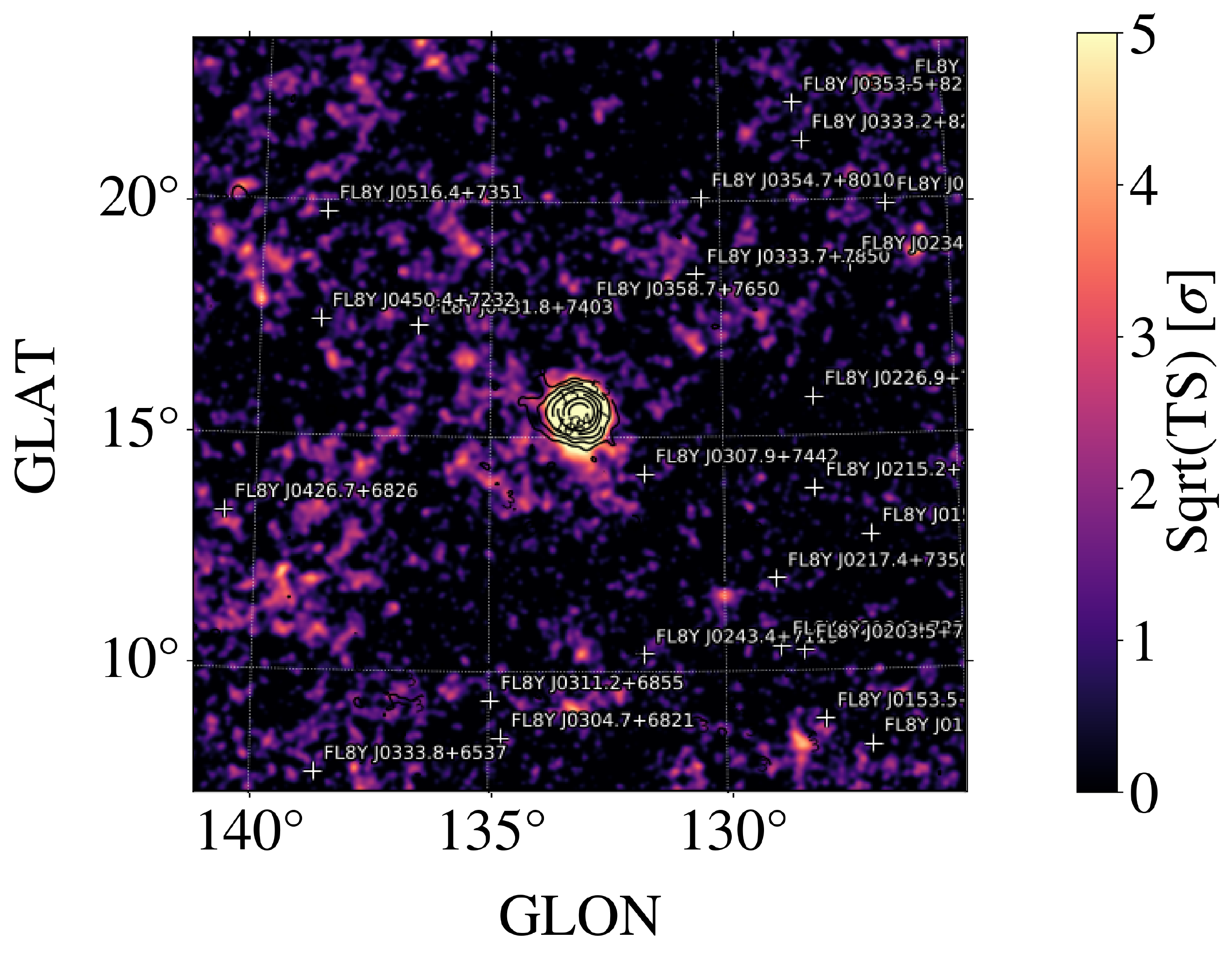}
\includegraphics[height=4.4cm]{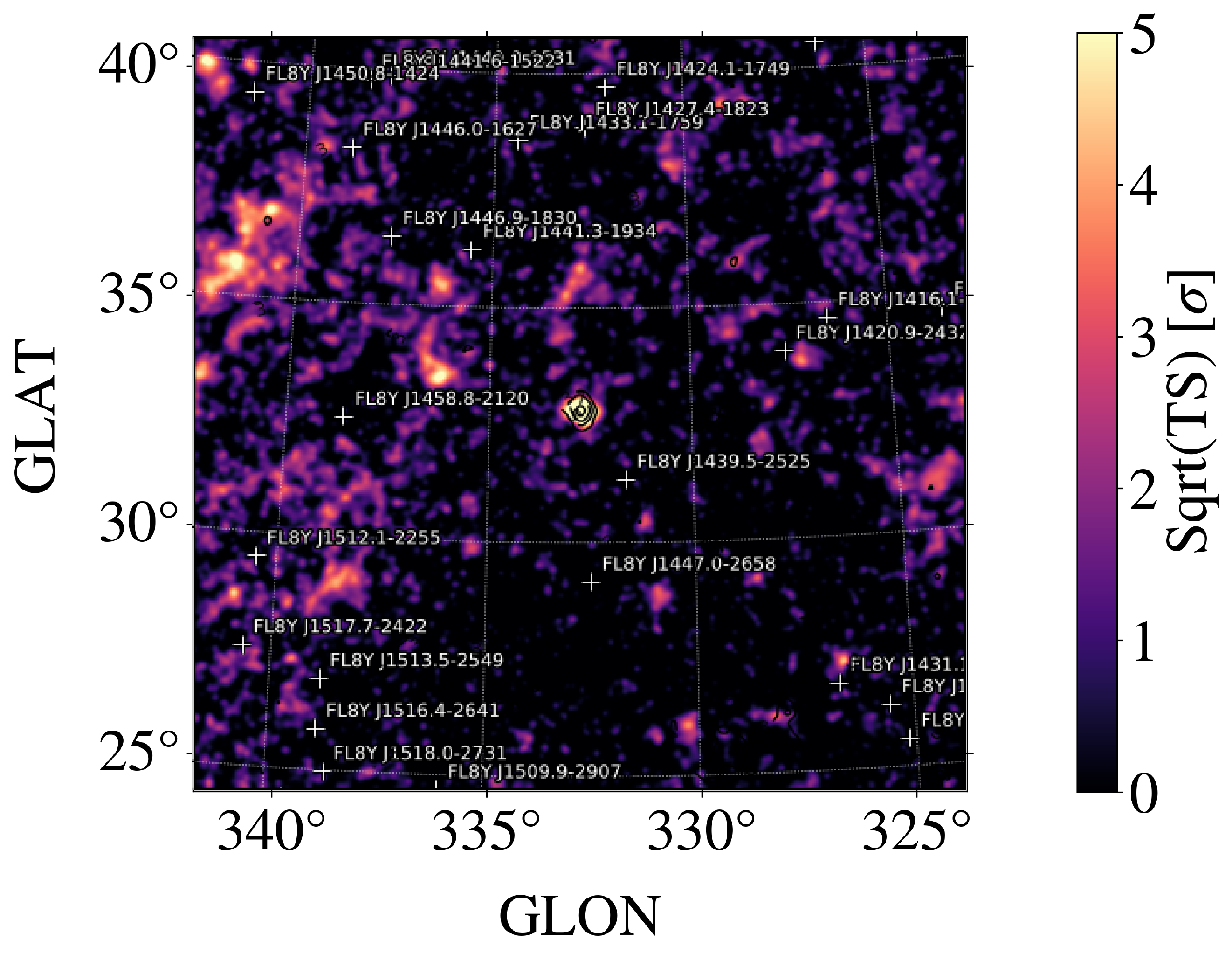}
\vfill
\includegraphics[height=4.4cm]{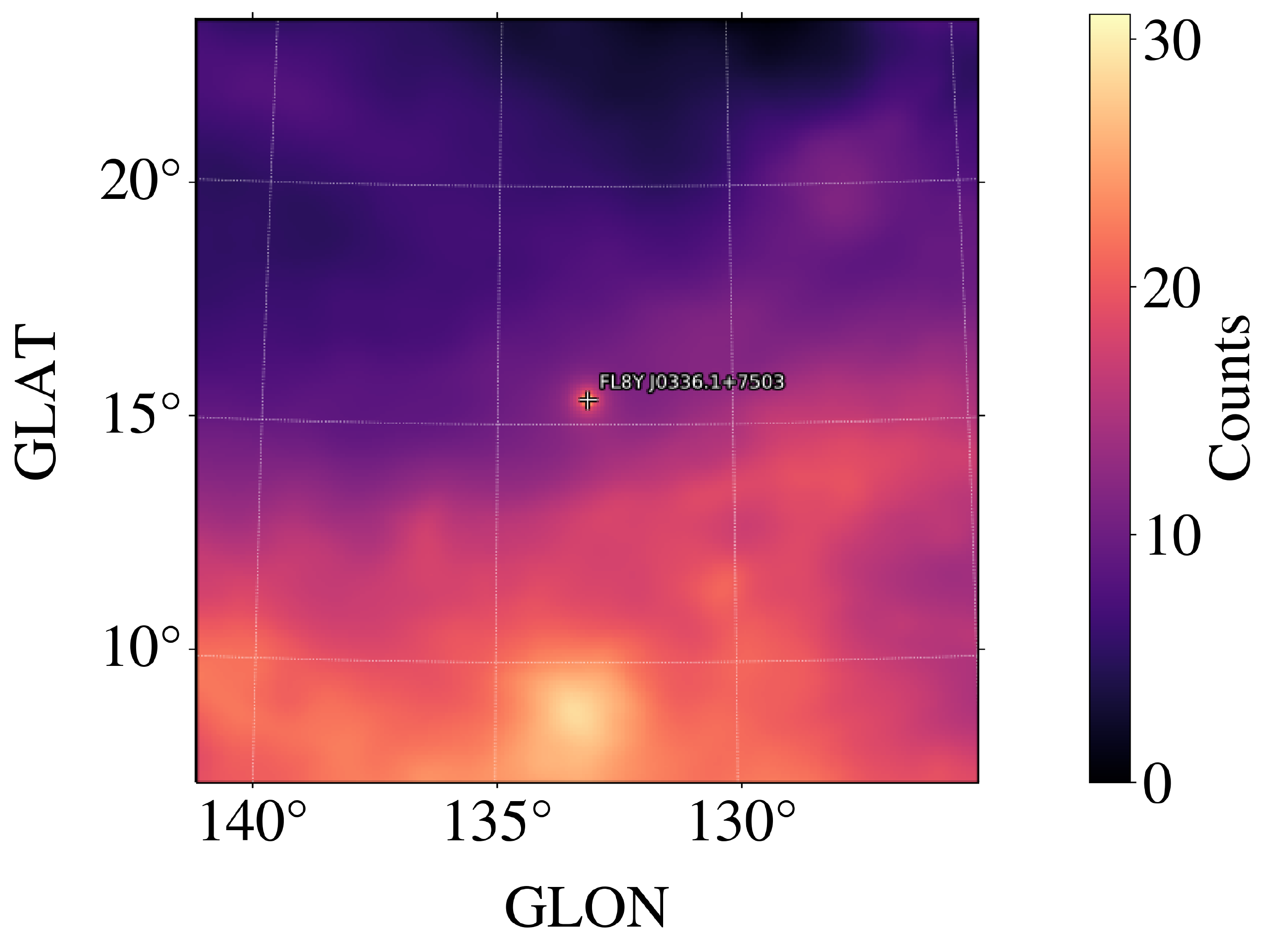}
\includegraphics[height=4.4cm]{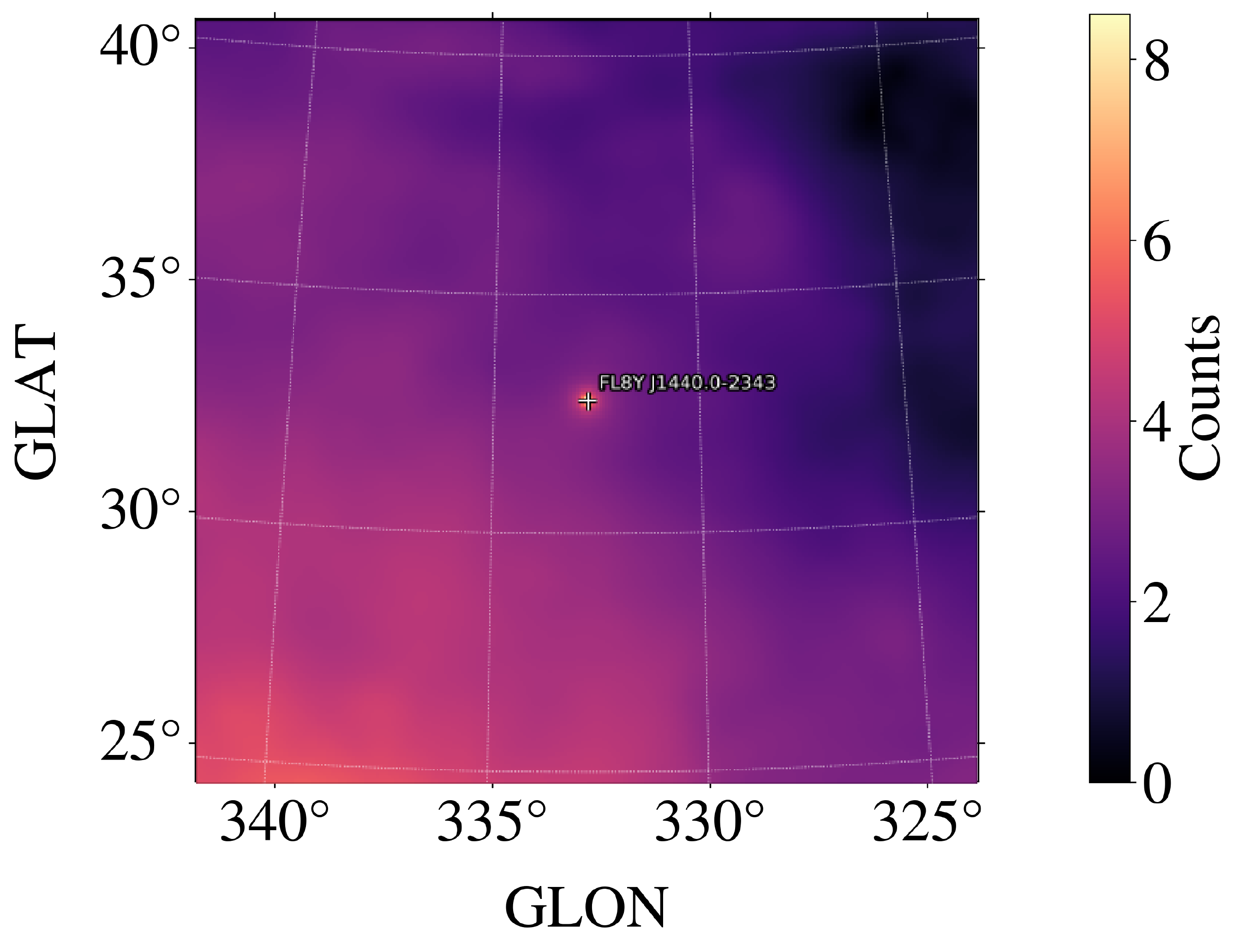}
\vfill
\includegraphics[height=4.4cm]{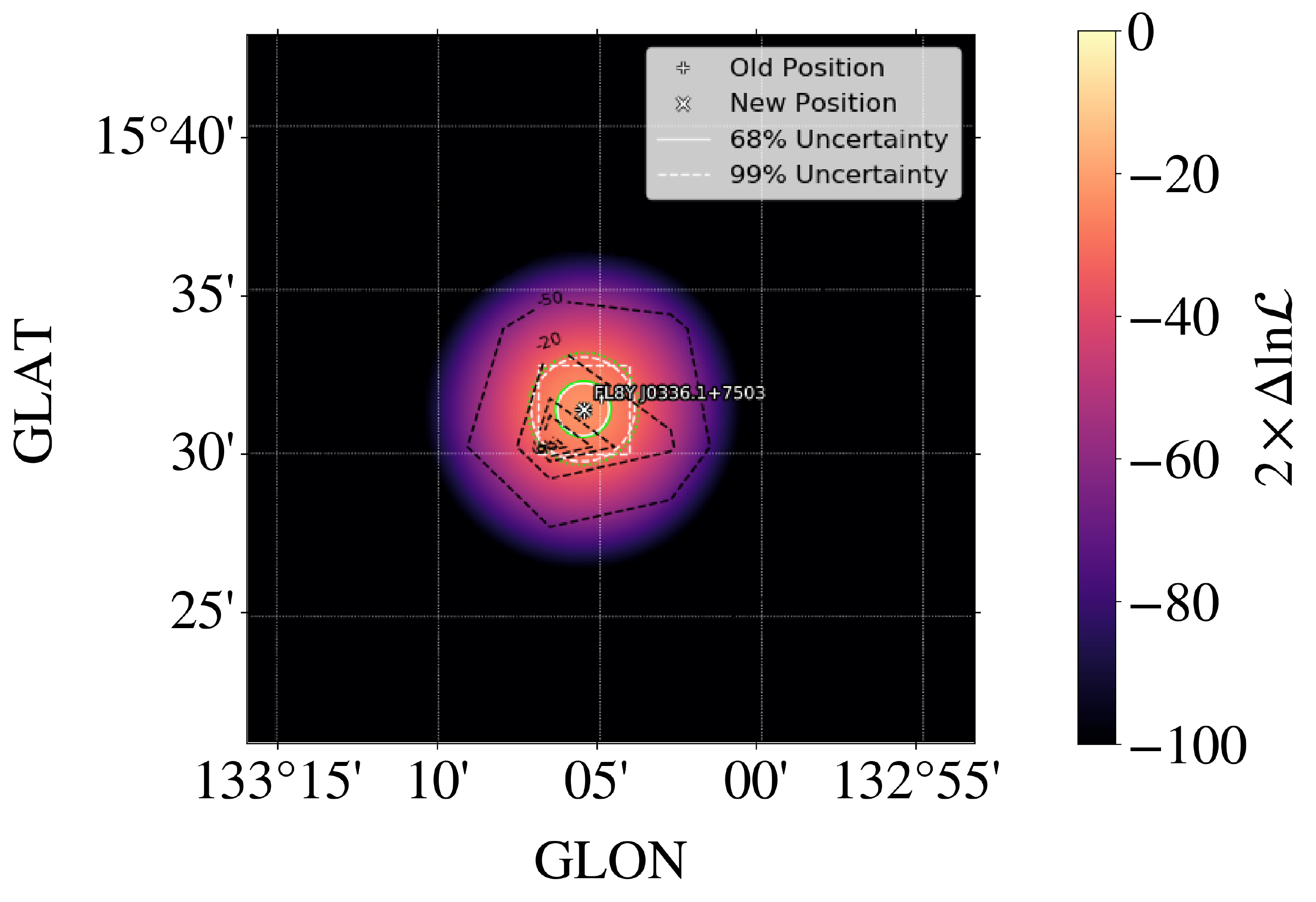}
\includegraphics[height=4.4cm]{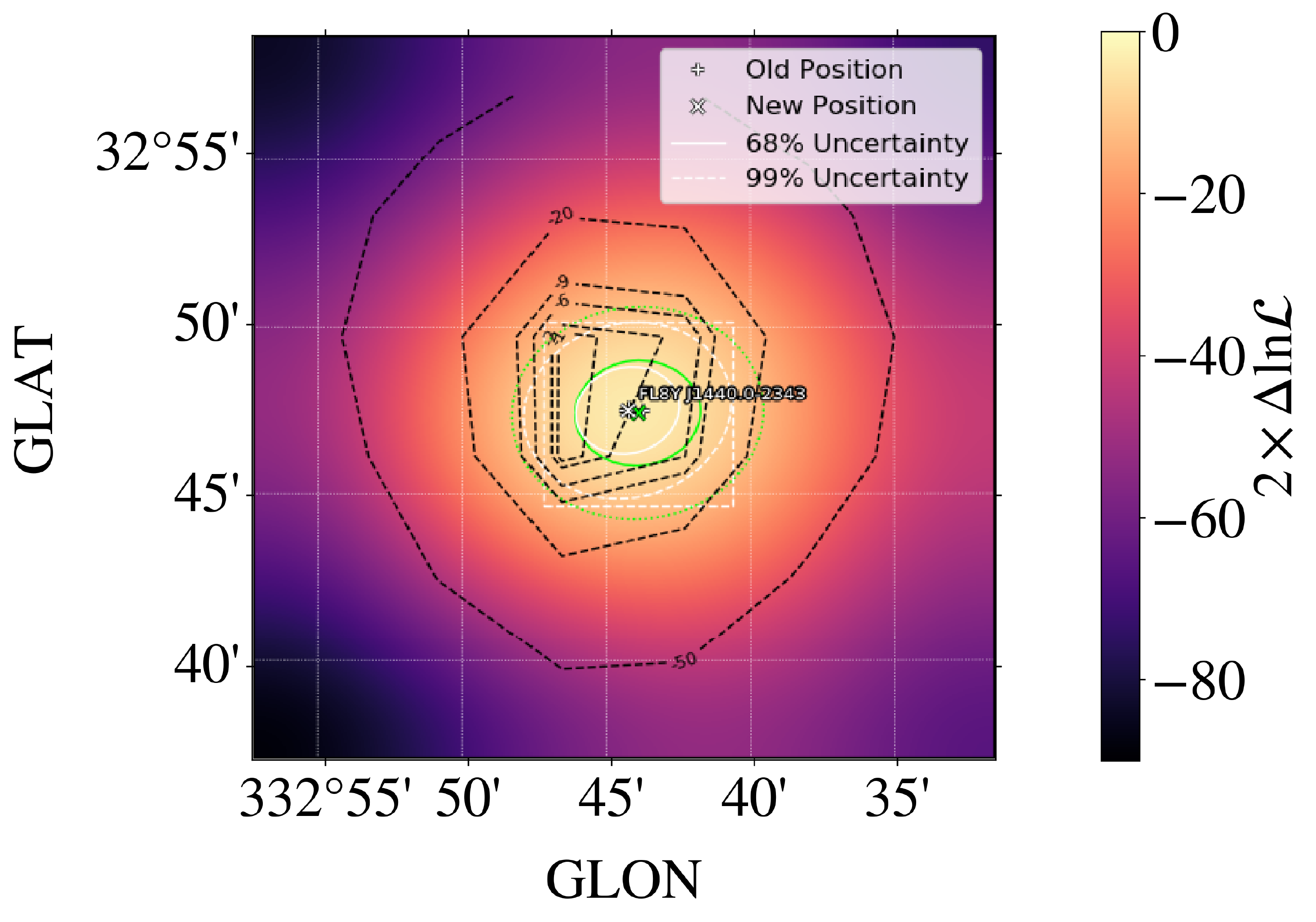}
\vfill
\includegraphics[height=4.7cm]{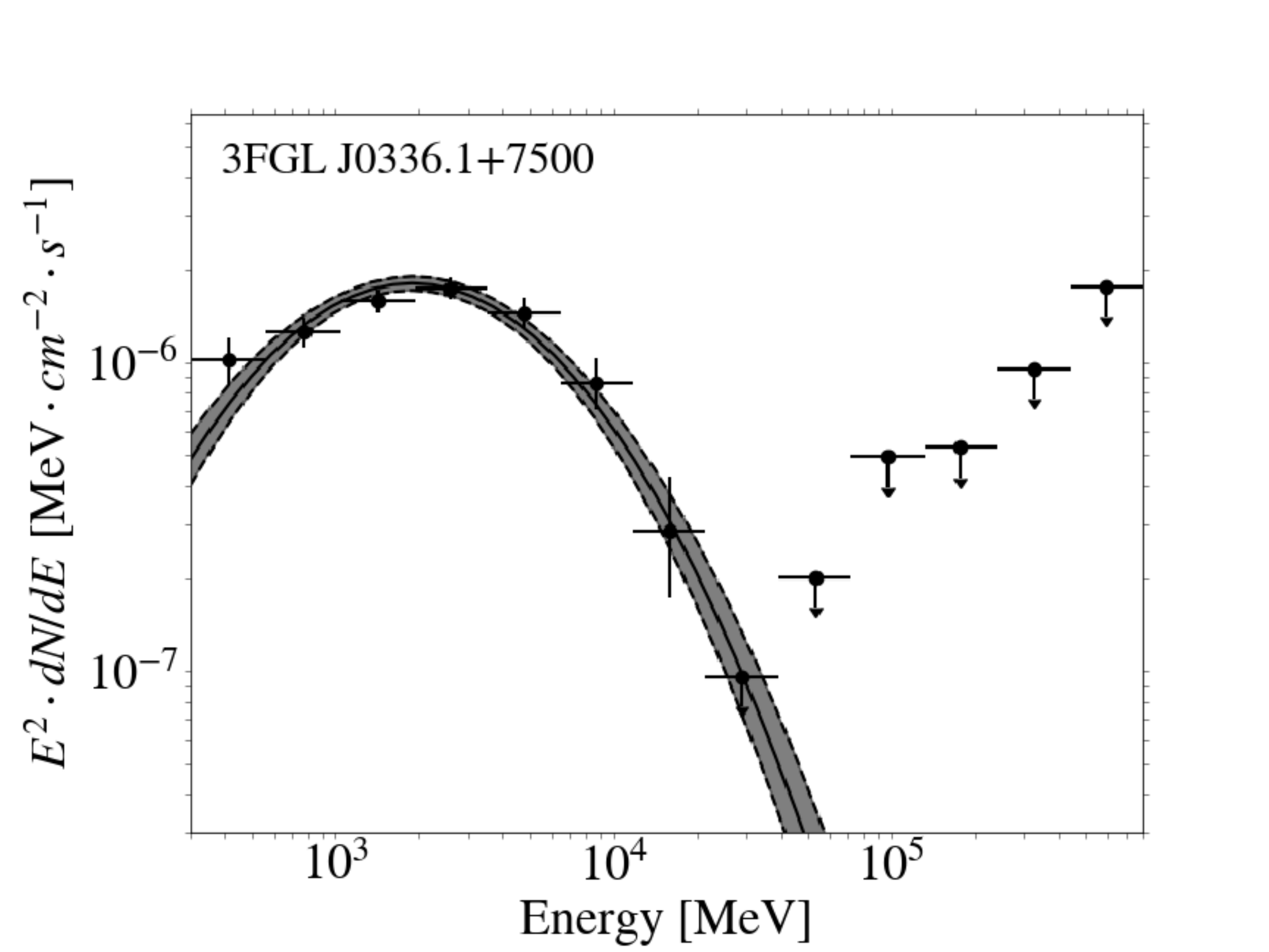}
\includegraphics[height=4.7cm]{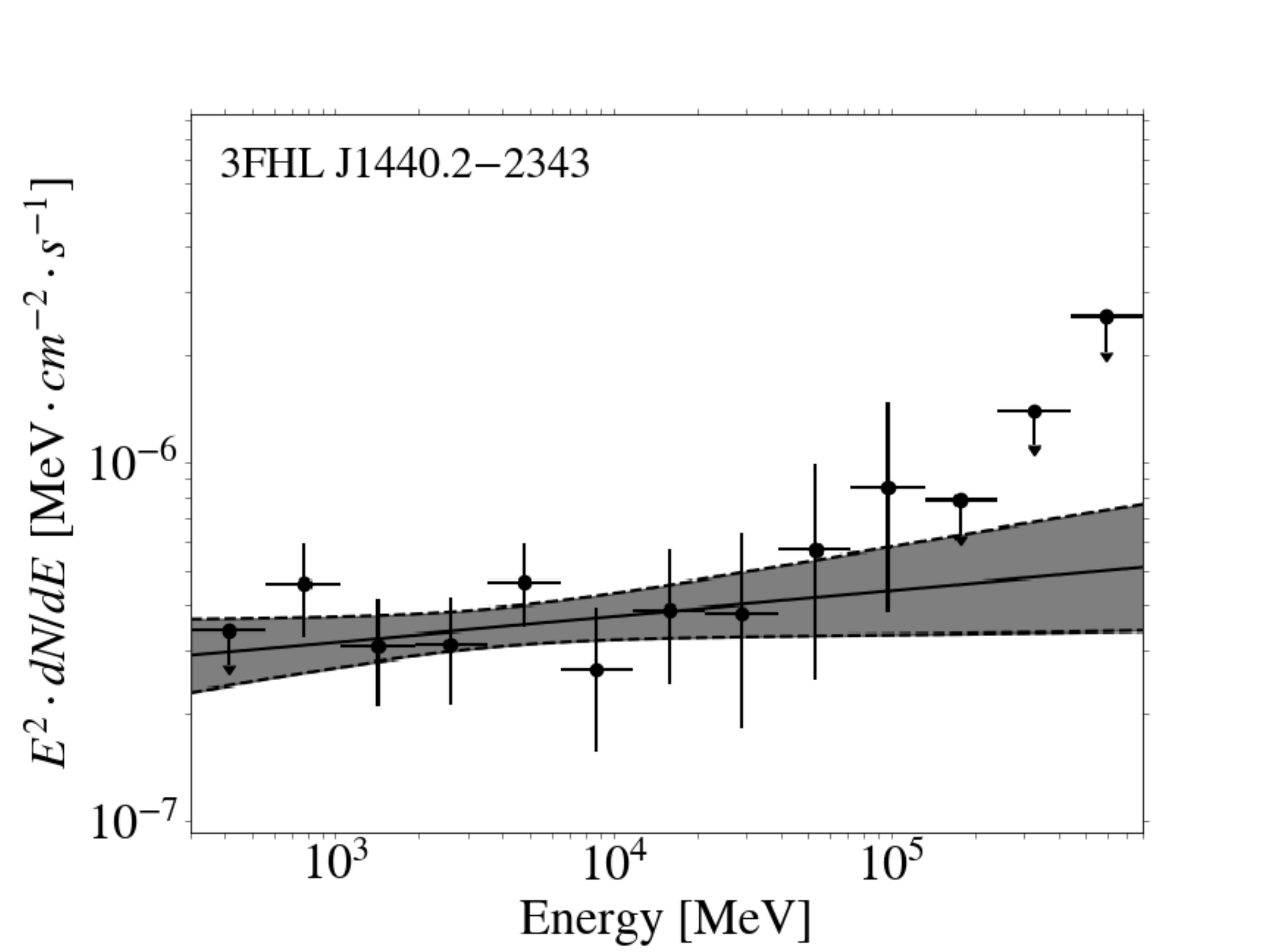}
\caption{Results examples of our spectral analysis, for two different types of sources in our unIDs' list: 3FGL J0336.1+7500 (left panels) and 3FHL J1440.2$-$2343 (right panels). The first is a bright source (detected at TS=1095) whose SED is best described by a PLE (see Eq. \ref{eq:powerlaw_exponentialcutoff}); while the second is a faint (TS=135) object for which a PL (Eq. \ref{eq:powerlaw}) is preferred. First row is the ROI residual map leaving the unID unmodelled, so that the TS peak is clearly visible. Second row shows the model of expected counts in the ROI. The location of the source, compared with the FL8Y reported position (green circle), is shown in the third row. The fourth row features the gamma-ray SED with the best-fit spectral model and $1\sigma$ uncertainties overimposed (solid line with gray shaded regions).}
\label{fig:analysis_example}
\end{figure}

Figure \ref{fig:analysis_example} shows the results of some of the steps in our analysis, for two different types of sources in our unIDs list, i.e., with SEDs well described by PL and PLE in each corresponding case.

From the DM perspective, this procedure assumes that the SED is entirely produced by DM annihilations. The fit is performed with \texttt{DMFitFunction} \cite{DMFit_paper}\footnote{We will later use the DM spectra as computed by \cite{Cirelli+12}, which only differs from this one for very heavy WIMPs, beyond the WIMP masses that this study probes.}. The differential spectrum for a pure annihilation channel\footnote{We focus on pure annihilation channels in order to stay agnostic to the underlying particle physics model.} is given by

\begin{equation}\label{eq:DMFit}
\frac{dN}{dE}=K\cdot F_{m_{\chi},C},
\end{equation}

\noindent where $K=\langle\sigma v\rangle\cdot J$ is the normalization, product of velocity-averaged annihilation cross section times the so-called J-factor, $F_{m_{\chi},C}$ is the photon flux for channel $C$ and DM particle mass $m_{\chi}$ at the energy under consideration. We fit the spectra to eight different annihilation channels, namely $b\bar{b}$, $\tau^+\tau^-$, $t\bar{t}$, $c\bar{c}$, $Z^0Z^0$, $W^+W^-$, $e^+e^-$ and $\mu^+\mu^-$. For each channel then, this DM fit will return the log-likelihood, the DM mass and $K$ with their $1\sigma$ uncertainties.

Once both the best-fit astrophysical and DM models have been obtained, our goal is to evaluate the statistical preference of DM against an astrophysical origin of the emission for each source in our DM subhalo candidates' list.

\subsection{Analysis results and DM hypothesis test}
\label{sec:spectral_results}
We perform the analysis of the 44 (4 2FHL, 24 3FHL and 16 3FGL) unID candidates that were found to be compatible with DM in \cite{Coronado_Blazquez2019}. The full list with all unIDs can be found in the supplementary material of \cite{Coronado_Blazquez2019}. Two out of these 44 unIDs, 3FHL J0350.4$-$5153 and 3FHL J1650.9+0430, are rejected as DM subhalos, as an X-ray counterpart was found in a multiwavelength search with \textit{Swift}-XRT \cite{Kaur2019}. An additional source, 3FHL J0233.0+3742, is also removed as it is associated as an active galaxy of uncertain type (BCU) in the 4FGL catalog \cite{4fgl_paper}.

We note that the detection significance of a true source is expected to increase as the square root of time. Yet, a total of nine sources (2FHL J1630.0+7644, 3FGL J0244.4+4745, 3FGL J0538.8$-$0341, 3FGL J1958.2$-$1413, 3FHL J0055.8+4507, 3FHL J0121.8+3808, 3FHL J0550.9+5657, 3FHL J1403.4+4319 and 3FHL J1726.2$-$1710) lie now below the detection threshold ($TS=25$) and therefore are removed from our sample\footnote{The reason why they do not reach the threshold may be different in each case. For the 3FGL sources we use different diffuse and isotropic models, as well as Pass 8 events instead of Pass 7. For 2FHL and 3FHL a possible cause may be that these sources were previously detected in a flaring state, which is now over (though given their low detection significances, just above the threshold, a statistical fluctuation is actually more plausible than a past flaring state.}. We also found some sources for which the corresponding detection TS ($\mathrm{TS_d}$) did not change as expected when considering a larger integration time, indeed still remaining near the detection threshold. This may be a hint of statistical artifacts due to mis-modeling of the diffuse emission. 3FHL J0041.7$-$1608 ($\mathrm{TS_d}=29.48$), 3FHL J0115.4$-$2916 ($\mathrm{TS_d}=30.77$) and 3FHL J0838.5+4006 ($\mathrm{TS_d}=26.66$) show this behavior\footnote{In fact, the TS is found to hardly increase with respect to the TS in the published catalogs.}. Nevertheless, in the conservative spirit of this work, we do not discard these sources as potential DM subhalos, as doing it would improve the constraints.

For the remaining 32 unIDs, we test the DM hypothesis against the astrophysical origin of the emission. In order to compare the log-likelihoods of the best astrophysical model vs. DM, we must take into account the degrees of freedom (d.o.f.), $k$, of each model \cite{Wang2016}. Indeed, from Eqs. \ref{eq:powerlaw} to \ref{eq:DMFit}, it can be seen that PL has the same number of d.o.f. as \texttt{DMFit}, $k=3$, while LP has $k=4$ and PLE $k=5$. Therefore, those models with more d.o.f. might provide better fits in some cases just because they have more parameters.

To penalize those models with higher d.o.f, we use the Akaike Information Criterion (AIC) \cite{Akaike1974},

\begin{equation}
\label{eq:akaike}
\Delta\mathrm{TS}=2\cdot\left[\left(k_0-k_\chi\right)+\textrm{log}\left(\frac{\mathcal{L}_\chi}{\mathcal{L}_0}\right)\right]
\end{equation}

\noindent where $k_0$ and $k_\chi$ are the d.o.f. of the astrophysical and DM best-fit models, respectively, and $\mathcal{L}_\chi$ ($\mathcal{L}_0$) is the likelihood of the DM (astrophysical) best-fit model. 

\begin{figure}[!ht]
\centering
\includegraphics[height=5.3cm]{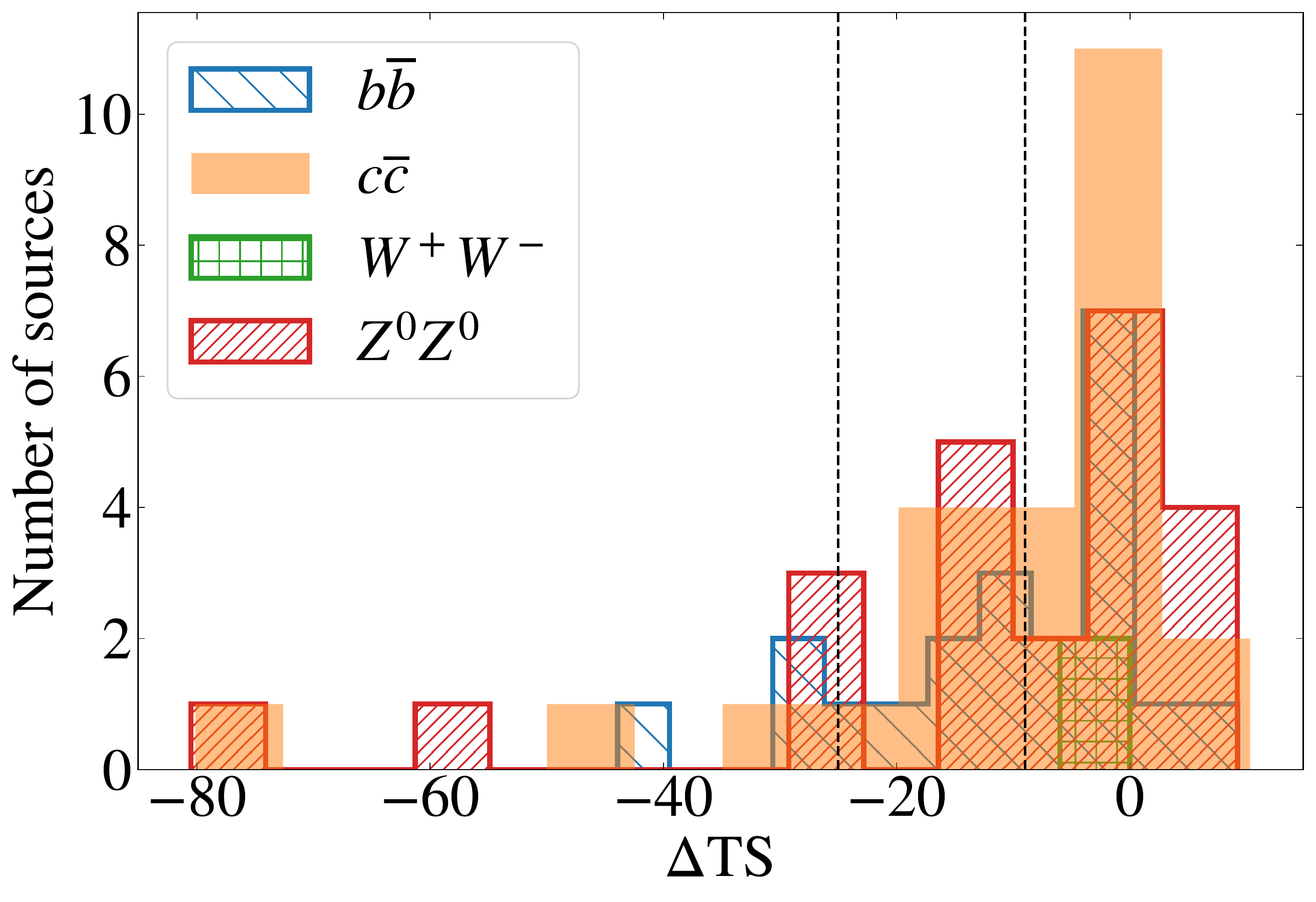}
\includegraphics[height=5.3cm]{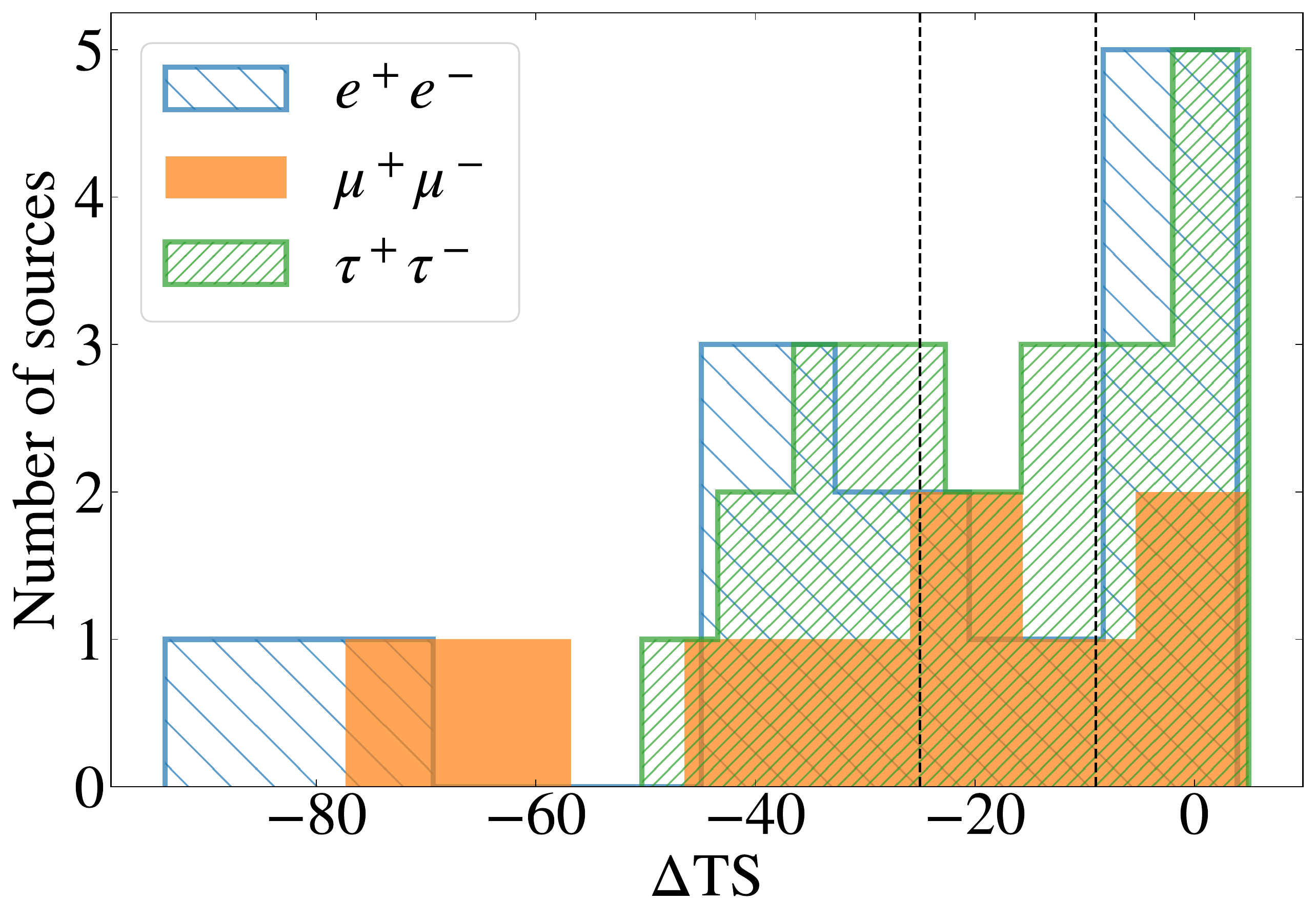}
\caption{Distribution of $\Delta$TS values as defined in Eq. \ref{eq:akaike} for the comparison between astrophysical models and DM fits. The left (right) panel is for hadronic (leptonic) annihilation channels. The two vertical dashed lines mark $\Delta TS=-25$ and $\Delta TS=-9$, i.e., the best-fit astrophysical model is preferred against DM at $\sim$5 and $\sim$3$\sigma$, respectively. Note that the $t\bar{t}$ channel does not appear in these plots, as no source is found to be well fitted by it.}
\label{fig:ts_hist}
\end{figure}

Figure \ref{fig:ts_hist} shows the $\Delta$TS distribution for all fitted sources and channels. The most successful annihilation channel is $b\bar{b}$, followed by $c\bar{c}$ and $\tau^+\tau^-$. Other leptonic channels such as $e^+e^-$ and $\mu^+\mu^-$ are able to fit only five and two sources, respectively. The less successful channels are $W^+W^-$, that can only fit two sources, and $t\bar{t}$, with no sources.

From our analysis, no unID in our initial list is found to prefer a DM origin significantly. Indeed, only seven sources have (at least) one annihilation channel marginally preferred over an astrophysical model ($\Delta$TS>0). However, all of them are far from being significant. Indeed, the maximum $\Delta$TS reached is 10.3, well below the $5\sigma$ significance threshold ($\Delta$TS$\sim$25) \cite{dsphs_paper}. Only in a couple of cases there is a $\sim$3$\sigma$ preference for DM. For a given channel, we will reject as DM subhalos those sources with $\Delta$TS>-9, i.e., the astrophysical model preferred at most at $\sim$3$\sigma$. This ensures the false-negative ratio to be negligible. In any case, we note that the obtained $\Delta$TS are pre-trials. Any potential factor that should be applied to correct for trials would go in the direction of decreasing the DM significance values. We do not perform such correction here, as the largest DM significances we found are already very low and thus the derivation of post-trials significances is not necessary.

\begin{figure}[!ht]
\centering
\includegraphics[height=5cm]{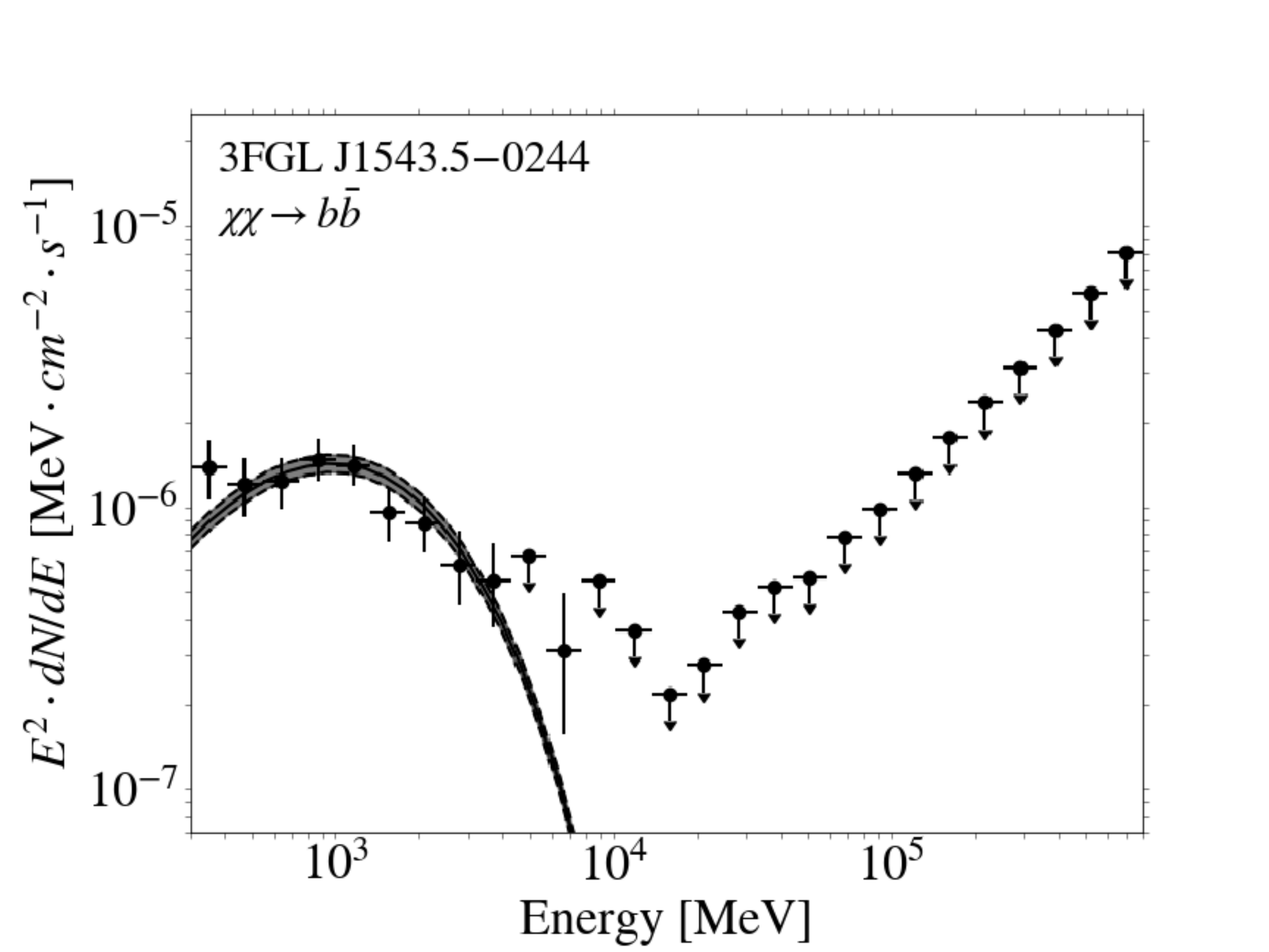}
\includegraphics[height=5cm]{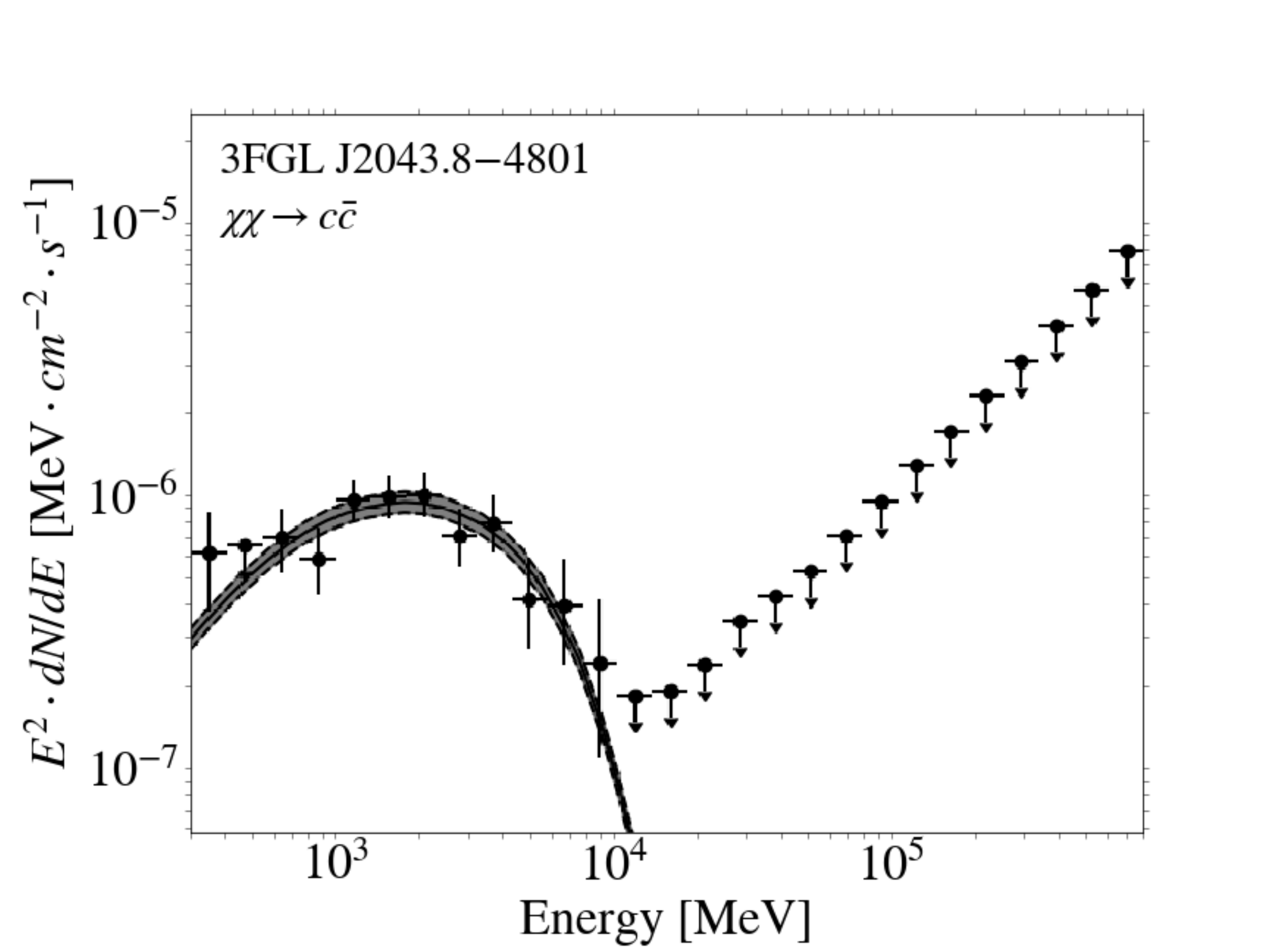}
\vfill
\includegraphics[height=5cm]{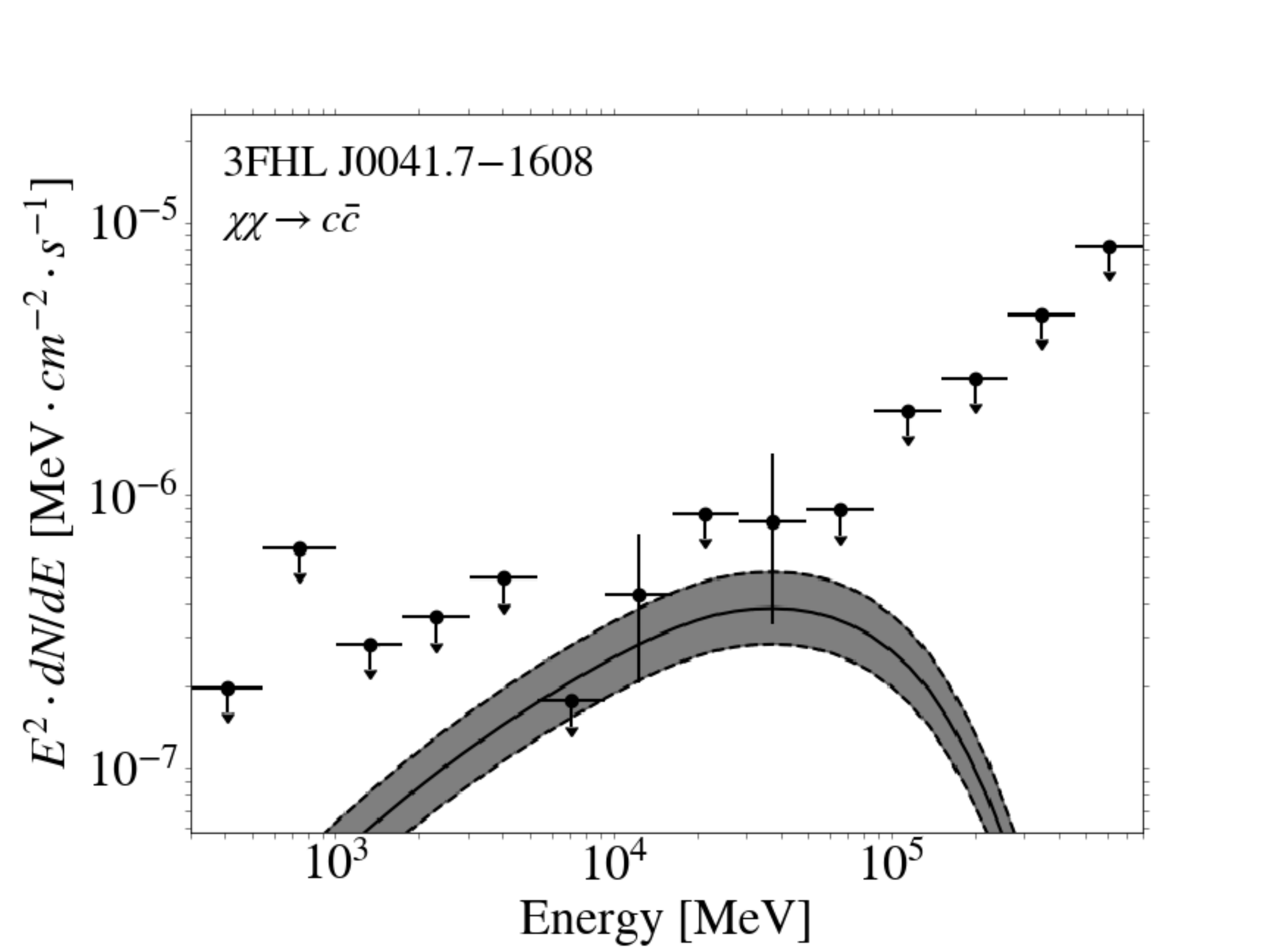}
\includegraphics[height=5cm]{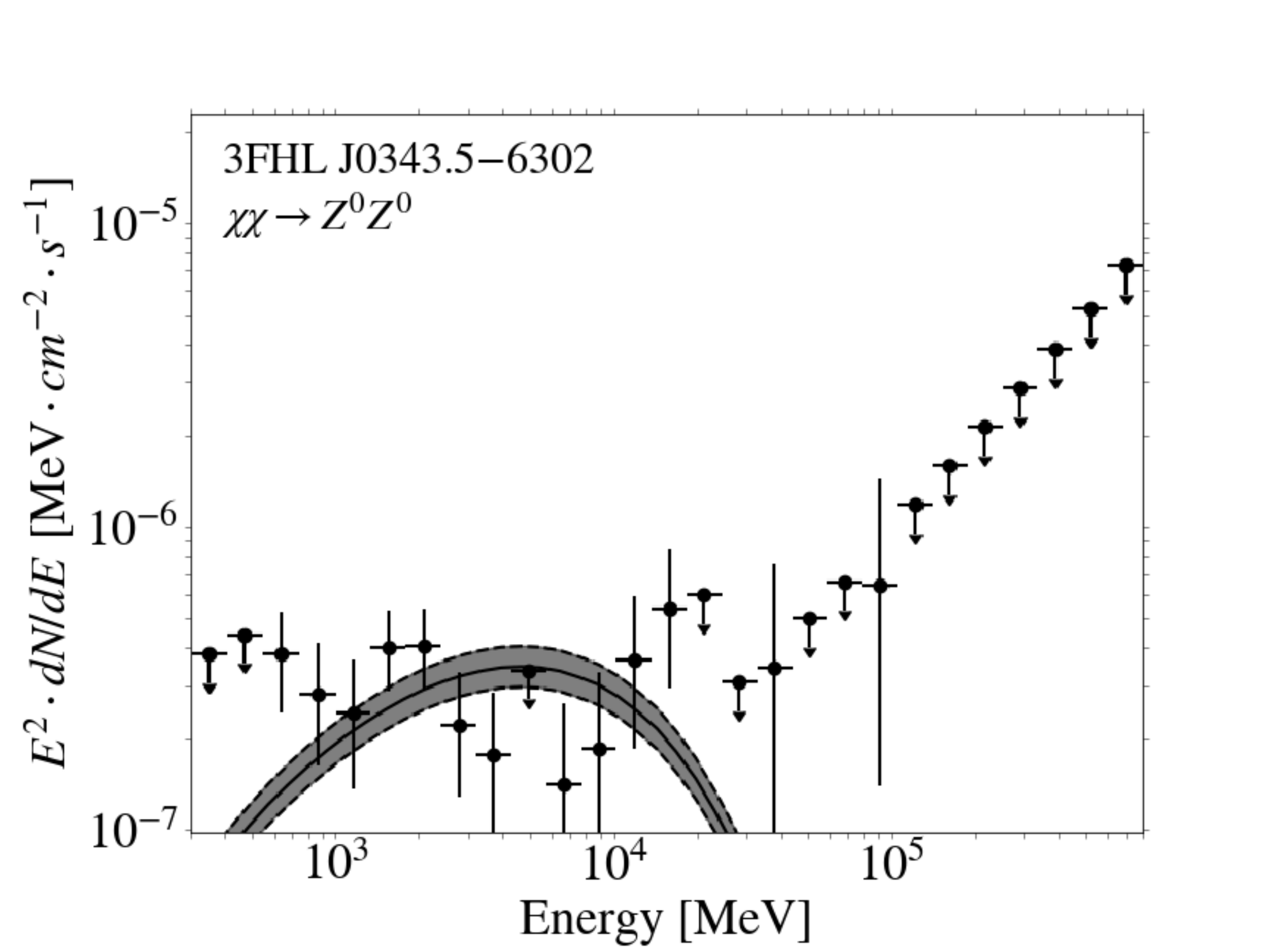}
\vfill
\includegraphics[height=5cm]{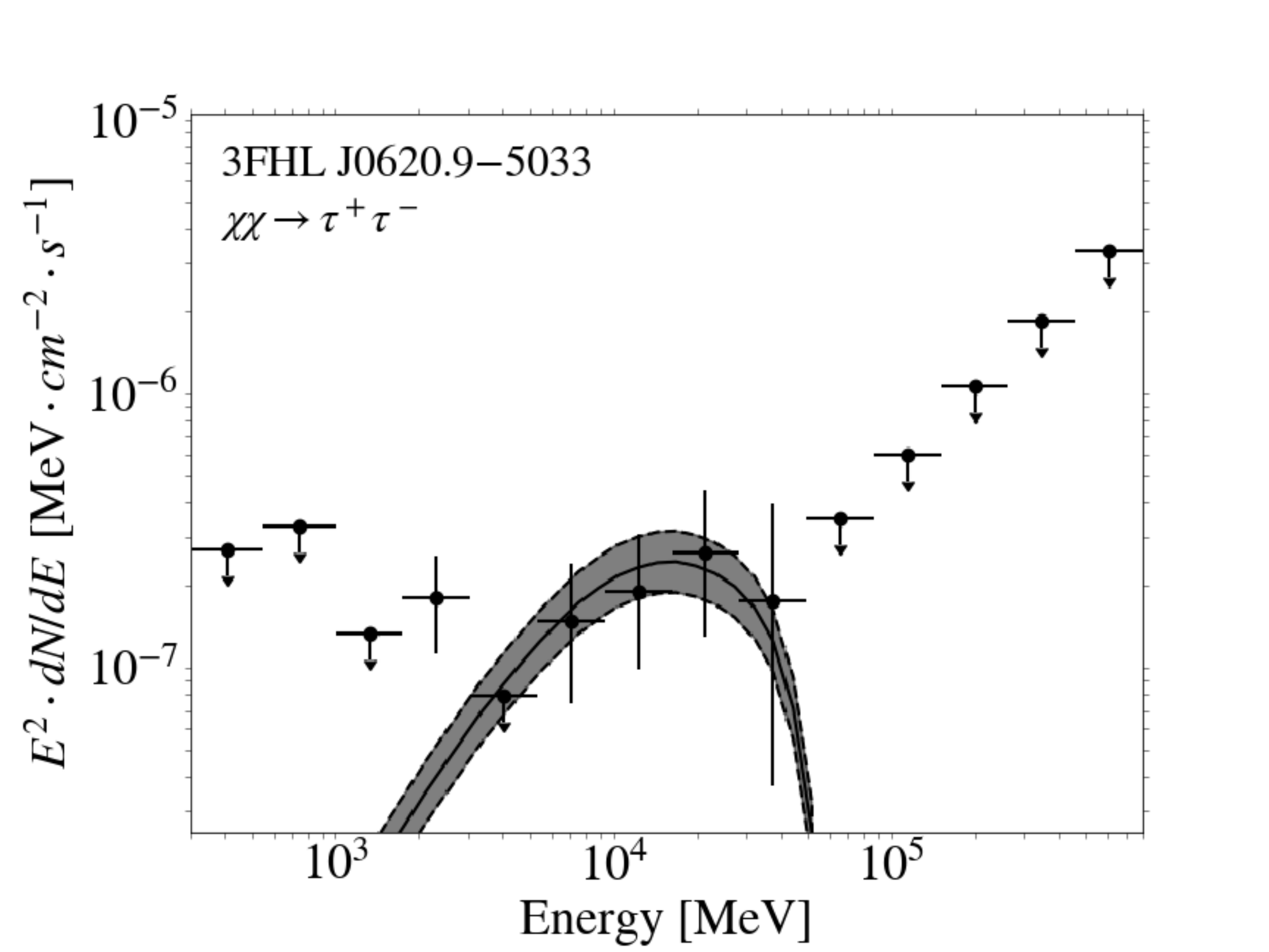}
\includegraphics[height=5cm]{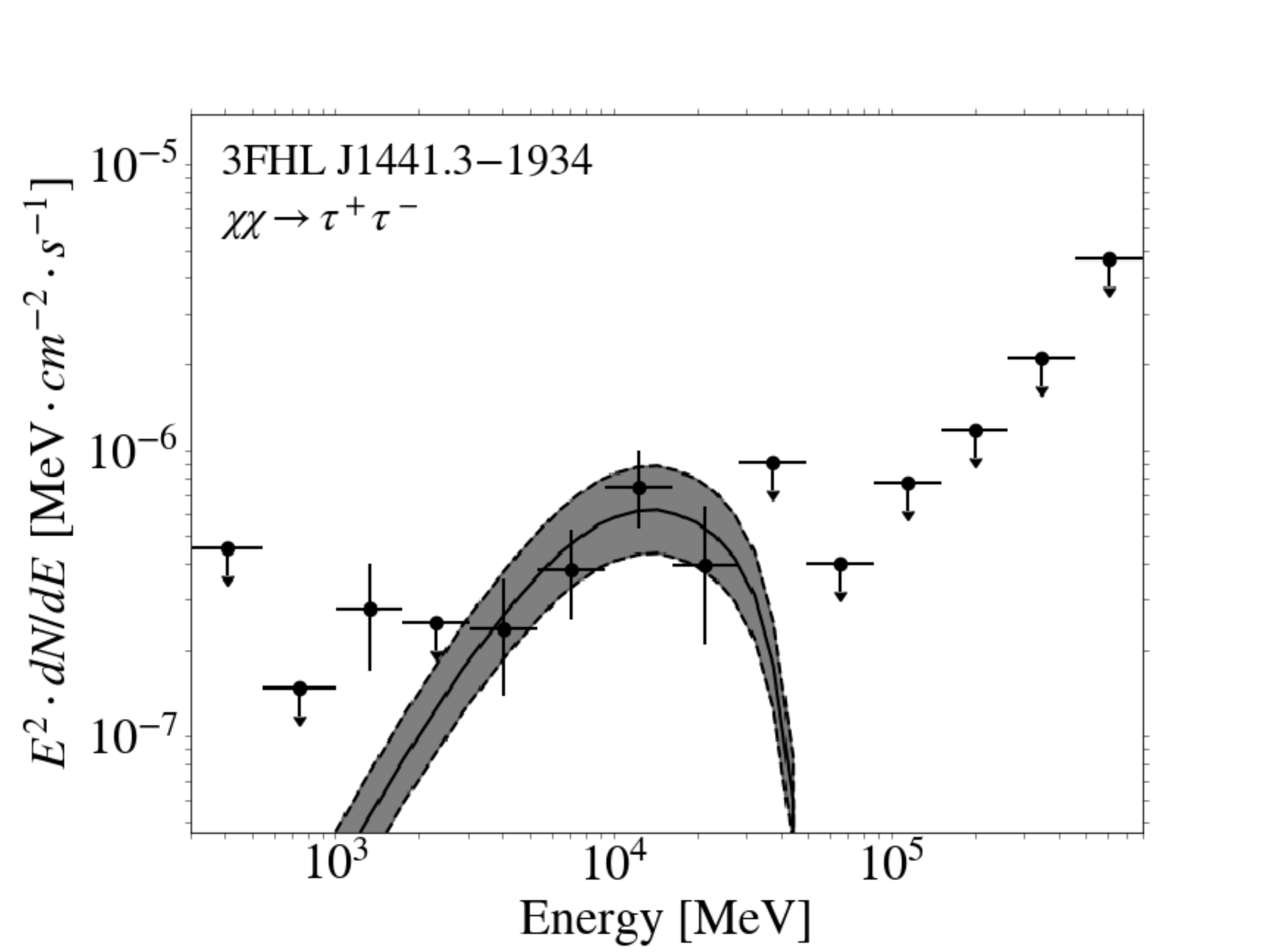}
\vfill
\includegraphics[height=5cm]{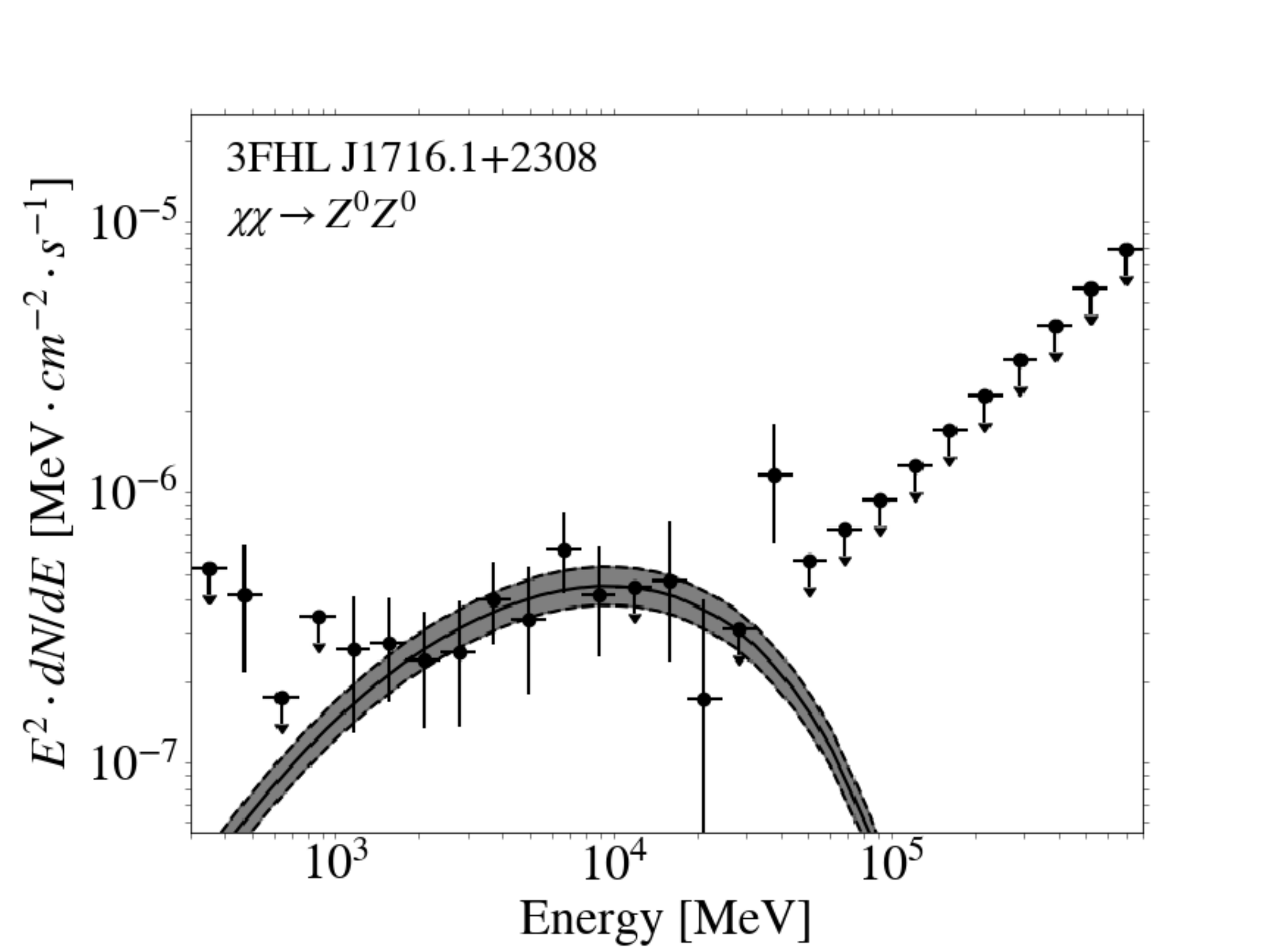}
\caption{SED of the seven candidates with $\Delta$TS>0 listed in Table \ref{tab:sources_summary}. Overimposed is the best-fit DM annihilation channel in each case. Note the significantly smaller model uncertainties in the case of the 3FGL sources (top panels).}
\label{fig:sources_dmfit}
\end{figure}

In Table \ref{tab:sources_summary}, we summarize the best fit DM models for those cases in which $\Delta$TS>0. The table provides the annihilation channel, the $\Delta$TS and the DM mass of the best fit. The SEDs of these seven objects, with the best-fit DM annihilation channel overimposed, are shown in Figure \ref{fig:sources_dmfit}. Note that the choice of number of bins is found to only marginally affect the $\Delta$TS. A full table with the selection criteria for all unIDs can be found in Appendix \ref{app:rejection}.

\begin{table}[!ht]
  \begin{center}
    \begin{tabular}{|M{3.6cm}|M{2.5cm}|M{1.3cm}|M{2.7cm}|M{2.6cm}|}
    \hline 
    Source name & DM channel(s) & $\Delta$TS & $m_\chi$ (GeV) & Astro. models\\
    \hline
    \hline
    3FGL J1543.5$-$0244& \makecell{$b\bar{b}$\\$Z^0Z^0$\\$c\bar{c}$} & \makecell{9.26\\ 8.29\\6.27} & \makecell{$15.2\pm1.3$\\$11.1\pm0.9$ \\$11.8\pm0.7$} & LP, PLE \\
    \hline
    3FGL J2043.8$-$4801 & \makecell{$c\bar{c}$\\$Z^0Z^0$\\$\tau^+\tau^-$} & \makecell{10.31\\9.21\\3.21} & \makecell{$22.4\pm1.7$\\$23.3\pm5.7$\\$8.5\pm0.4$} & PLE, LP\\
    \hline
    3FHL J0041.7$-$1608 & {$c\bar{c}$} & 1.87 & $666\pm99$ & PL\\
    \hline
    3FHL J0343.5$-$6302 & \makecell{$Z^0Z^0$\\ $c\bar{c}$} & \makecell{3.89\\0.59} & \makecell{$112\pm14$\\ $67.1\pm6.7$} & PL, LP\\
    \hline
    3FHL J0620.9$-$5033 & $\tau^+\tau^-$ & 0.63 & $56.7\pm9.2$ & PL, PLE\\
    \hline
    3FHL J1441.3$-$1934 & \makecell{$\tau^+\tau^-$\\ $\mu^+\mu^-$\\ $e^+e^-$\\ $b\bar{b}$\\ $c\bar{c}$\\ $Z^0Z^0$} & \makecell{4.92\\4.91\\3.89\\2.79\\2.68\\2.12} & \makecell{$48.1\pm13.3$\\ $29.6\pm2.9$\\ $29.6\pm3.0$\\ $328\pm45$\\ $197\pm22$\\ $299\pm32$} & PLE, PL\\
    \hline
    3FHL J1716.1+2308 & \makecell{$Z^0Z^0$\\$c\bar{c}$} & \makecell{3.19\\2.59} & \makecell{$207\pm25$\\$162\pm47$} & PLE, PL\\
    \hline
    \end{tabular}
    \caption{Sources exhibiting positive $\Delta$TS values, sorted by catalog and right ascension (RA), following its definition in Eq. \ref{eq:akaike}, i.e., in these cases DM is preferred against astrophysical models. The second column shows the DM channel(s) for which we have positive $\Delta$TS. For almost all these sources, there are different channels providing $\Delta$TS>0. The third column gives their corresponding $\Delta$TS. These values are far from being significant; see also discussion in the text. The fourth column shows best-fit DM mass with $1\sigma$ uncertainty. The last column reports the best-fit astrophysical models.}
    \label{tab:sources_summary}
  \end{center}
\end{table}

\section{Limits on the DM annihilation cross section}
\label{sec:constraints}
Our analysis of unID spectral properties in \cref{sec:spectral_section} has been useful to safely decrease the original number of DM subhalo candidates. In the absence of outstanding DM candidates in the remaining list (see, e.g., the $\Delta$TS values in Table  \ref{tab:sources_summary}), we now proceed to set constraints on the DM annihilation cross section.

The procedure is the same as in \cite{Coronado_Blazquez2019}, and, thus, we refer the reader to the mentioned work for further details. The methodology consists on a comparison between the predicted number of subhalos in the simulations that should outshine in gamma rays for the LAT, and the actual number of remaining unIDs in LAT catalogs that may still be potential DM subhalos according to our studies. The smaller the number of remaining unIDs the stronger the DM constraints. In this sense, the constraints in this section represent an updated version of the constraints presented in \cite{Coronado_Blazquez2019}. Following our previous work, the velocity-averaged annihilation cross section can be expressed as:

\begin{equation}
\label{eq:master_formula}
\langle\sigma v\rangle=\frac{8\cdot\pi\cdot m_{\chi}^2\cdot F_{min}}{J\cdot N_{\gamma}}
\end{equation}

\noindent where $F_{min}$ is the instrumental sensitivity to DM, $J$ the nominal value of the subhalos' J-factor, and $N_{\gamma}$ the DM spectrum for a particular annihilation channel integrated within the energy range under consideration. In the current work, the constraining power depends on the number of sources that can be fitted by each annihilation channel, e.g., while for $b\bar{b}$ we have five 3FGL and six 3FHL sources that cannot be ruled out as DM subhalos, for $\tau^+\tau^-$ we have just one 3FGL and four 3FHL unIDs. There are not 2FHL sources in any channel.

\begin{table}
  \begin{center}
    \begin{tabular}{|M{1.2cm}|M{1.2cm}|M{1.2cm}|M{1.2cm}|M{1.2cm}|M{1.2cm}|M{1.2cm}|M{1.2cm}|M{1.2cm}|}
    \hline
     & $b\bar{b}$ & $c\bar{c}$ & $t\bar{t}$ & $\tau^+\tau^-$ & $e^+e^-$ & $\mu^+\mu^-$ & $W^+W^-$ & $Z^0Z^0$\\
    \hline
    \hline
    2FHL & 0 & 0 & 0 & 0 & 0 & 0 & 0 & 0\\
    \hline
    3FHL & 6 & 8 & 0 & 4 & 4 & 2 & 1 & 6\\
    \hline
    3FGL & 5 & 7 & 0 & 1 & 1 & 0 & 1 & 6\\
    \hline
    \end{tabular}
    \caption{Summary of the number of sources from the three different catalogs which are fitted to DM annihilation channels with $\Delta\mathrm{TS}\geq-9$. There is no overlap between catalogs, i.e., no source from a particular catalog is present in one of the other two. There is no 2FHL source or $t\bar{t}$ channel fit present in the final result.}
    \label{tab:remaining_sources}
  \end{center}
\end{table}

We conservatively consider as potential DM subhalos those unIDs for which the $\Delta$TS as defined in Eq. \ref{eq:akaike}, and obtained from spectral fits to DM and astrophysical models, is $\Delta TS\geq -9$, i.e., the spectral astrophysical models are preferred against DM at less than $3\sigma$. The weakest limits are for $\tau^+\tau^-$, with one 3FGL and four 3FHL sources and $c\bar{c}$, with seven sources both in 3FGL and 3FHL (see Table \ref{tab:remaining_sources} for all channels and catalogs). The strongest limits are for the $t\bar{t}$ channel, as it does not provide good fits to any source\footnote{As stated in \cite{Coronado_Blazquez2019}, we are in the sensitivity reach scenario.}. No DM subhalo annihilating into pure channels is present in the 2FHL catalog according to our spectral analysis, and therefore the corresponding DM constraints for this catalog are already at the level of sensitivity reach (i.e., the maximum potential of the method) for all considered channels.

As in \cite{Coronado_Blazquez2019}, and in order to obtain 95\% C.L. upper limits for a given number $n$ of remaining unIDs, we integrate the distribution of J-factor values corresponding to the $n^{th}$ brightest subhalo as obtained from our 1000 VL-II realizations, and take the J-factor above which 95\% of the whole distribution is contained as the value of $J$ to be used in Eq. \ref{eq:master_formula}.

\begin{figure}[!ht]
\centering
\includegraphics[height=8.5cm]{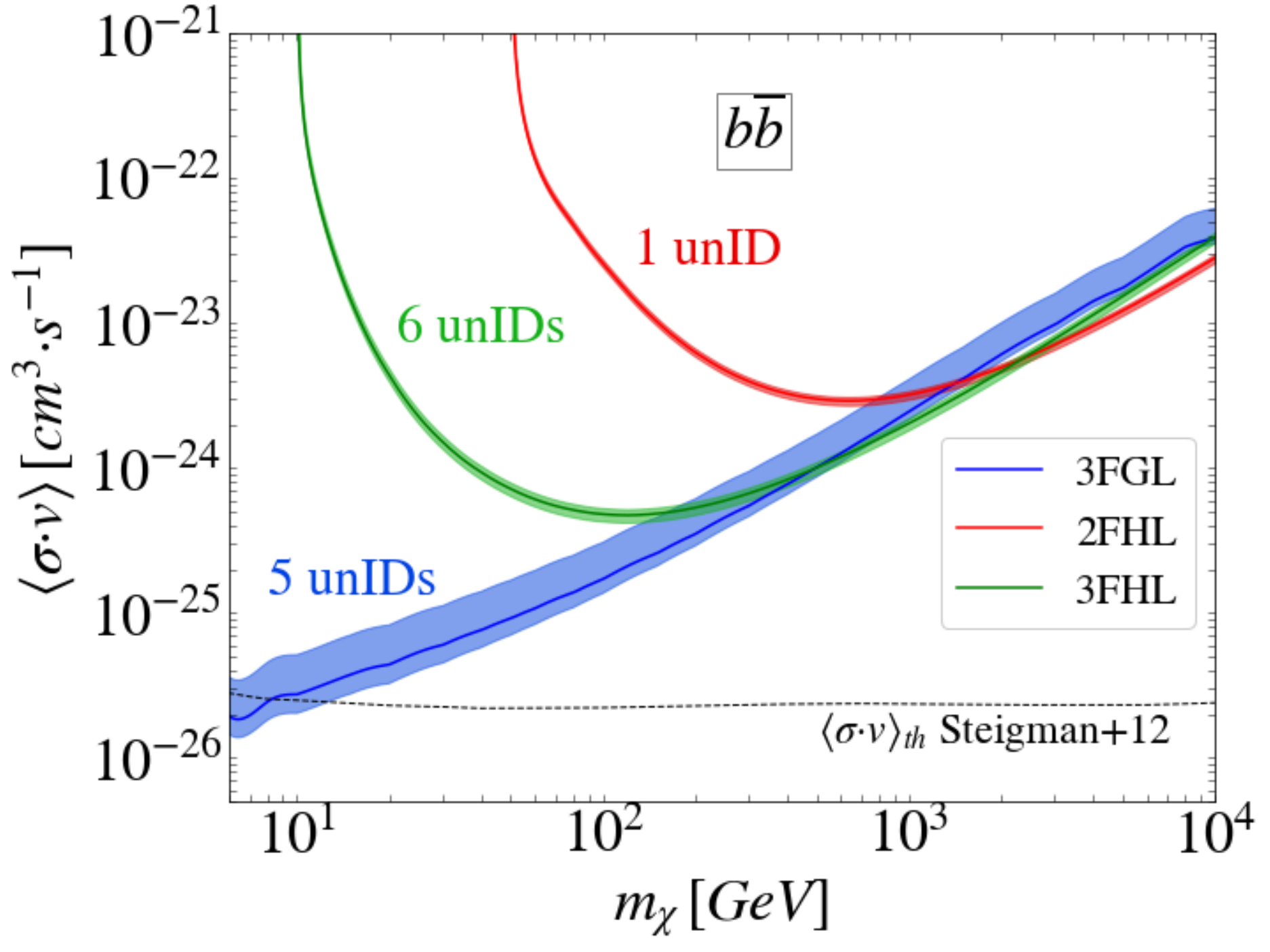}
\vfill
\includegraphics[height=8.5cm]{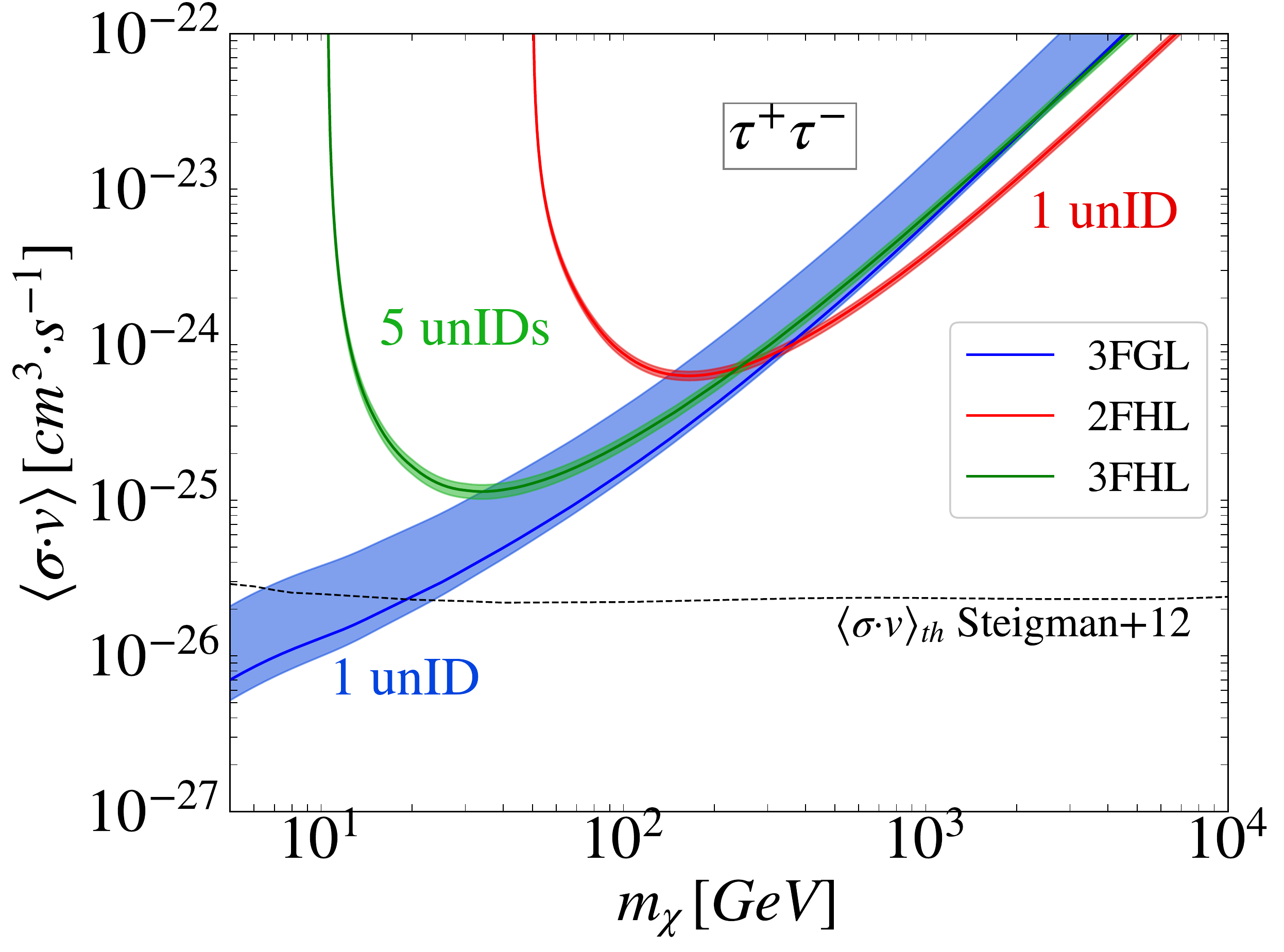}
\caption{Limits on the DM annihilation cross section for $b\bar{b}$ (top) and $\tau^+\tau^-$ (bottom) for the 3FGL, 2FHL and 3FHL LAT catalogs and the number of unIDs that are still compatible with a DM annihilation origin after our spectral analyses of \cref{sec:spectral_section} and \cref{sec:beta}. The wide uncertainty band in the case of the 3FGL constraint is due to false positive rates in the machine learning classification algorithms used in the unID filtering work of Ref. \cite{Coronado_Blazquez2019}. The dashed line represents the thermal value of the annihilation cross section \cite{Steigman+12}.}
\label{fig:limits_bb_tautau}
\end{figure}

In Figure \ref{fig:limits_bb_tautau} we present the constraints for $b\bar{b}$ and $\tau^+\tau^-$ annihilation channels. They are able to rule out canonical thermal WIMPs with masses up to $\sim$10 GeV for $b\bar{b}$ and $\sim$20 GeV for $\tau^+\tau^-$. With respect to our previous work, these constraints improve between a factor 2-5 depending on the annihilation channel. The uncertainty bands correspond to the $1\sigma$ uncertainty obtained after averaging $F_{min}$ over the whole sky. Also, in the case of the 3FGL constraint, we note that the upper half of the uncertainty band is larger than the lower half due to the way the machine learning conservative false positive rate is implemented in the unID filtering of \cite{Coronado_Blazquez2019}\footnote{In short, the expected false positive ratio ($\leq4\%$) is taken into account defining as the upper band the limit obtained considering a J-factor corresponding to six more sources (4\% of the used sample) remaining.}. As done in our previous work, in those cases where no unID is compatible with DM, we prefer to stay conservative and still adopt the J-factor of the brightest subhalo in the simulation. For the limits in Figure \ref{fig:limits_bb_tautau}, this is the case of the ones derived from the 2FHL catalog for $b\bar{b}$, and $\tau^+\tau^-$.

\begin{figure}[!ht]
\centering
\includegraphics[height=8.5cm]{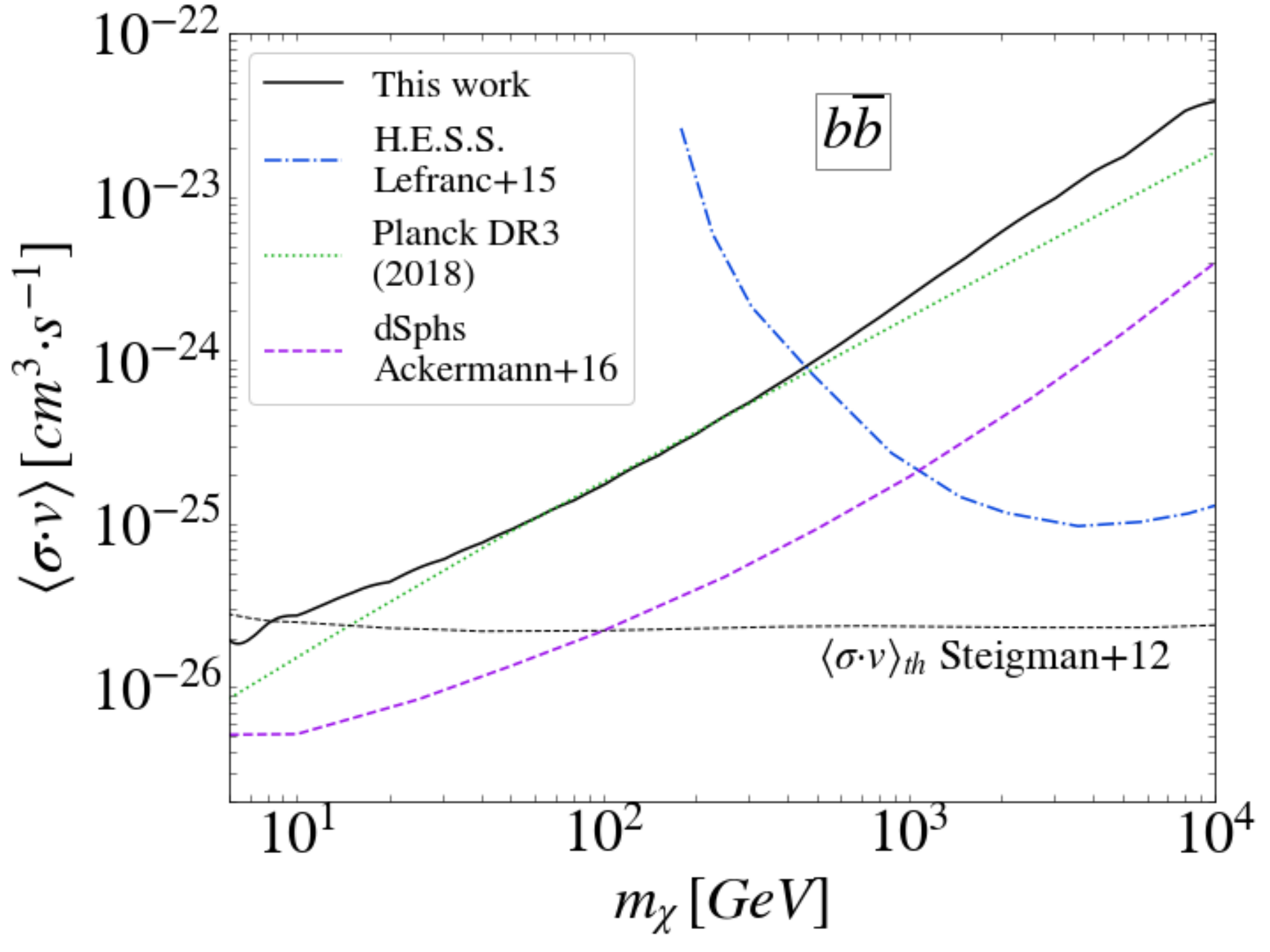}
\vfill
\includegraphics[height=8.5cm]{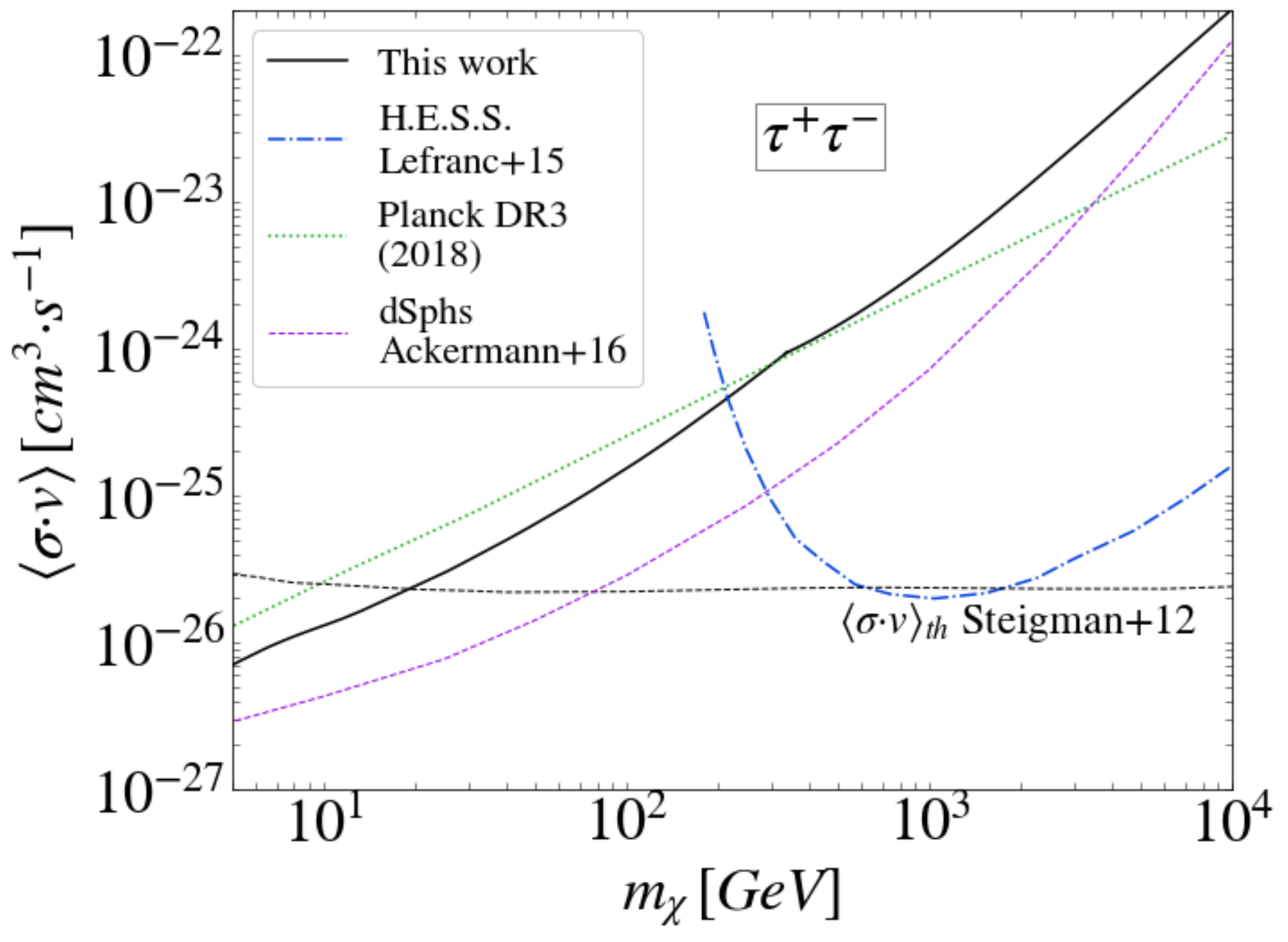}
\caption{Comparison of DM limits from different targets and probes. Solid line corresponds to the envelope of the three sets of limits shown in Figure \ref{fig:limits_bb_tautau}. Dot-dashed, dashed and dotted lines are, respectively, the DM constraints derived by H.E.S.S. for the Milky Way halo \cite{Hess_Lefranc+15}, by the \textit{Fermi}-LAT for dSphs \cite{dsphs_paper}, and by Planck using the CMB DR3 latest release \cite{Planck_dr3}. Top panel is for $b\overline{b}$ and bottom panel for $\tau^+\tau^-$ annihilation channel.}
\label{fig:envolvente_bb_tautau}
\end{figure}

Figure \ref{fig:envolvente_bb_tautau} shows the envelope of the limits presented in Figure \ref{fig:limits_bb_tautau}, together with other independent limits from targets such as the MW halo with H.E.S.S. \cite{Hess_Lefranc+15}, the cosmic microwave background with \textit{Planck} DR3 \cite{Planck_dr3} and the dSphs with \textit{Fermi}-LAT \cite{dsphs_paper}. Our limits are a factor 2-4 weaker than the dSphs constraints, while they are more competitive than H.E.S.S. below $\sim200-400$ GeV (depending on the channel). Also, in the case of $\tau^+\tau^-$, they improve \textit{Planck} DR3 constraints by a factor $\sim2$ up to 300 GeV.

\section{The $\beta$ plot: a novel technique to address DM/pulsar confusion}
\label{sec:beta}

Spectral confusion between pulsars and DM annihilation signals has been a long issue in the quest for DM in gamma rays (see for example \cite{Mirabal2013,Mirabal2016,fermi_dm_satellites_paper}), especially for low, $\mathcal{O}\left(10\right)$ GeV DM masses, and when considering hadronic channels such as $b\bar{b}$. In this section we propose a new test that could help distinguish them, at least in some particular cases.

\begin{figure}[!ht]
\centering
\includegraphics[height=9cm]{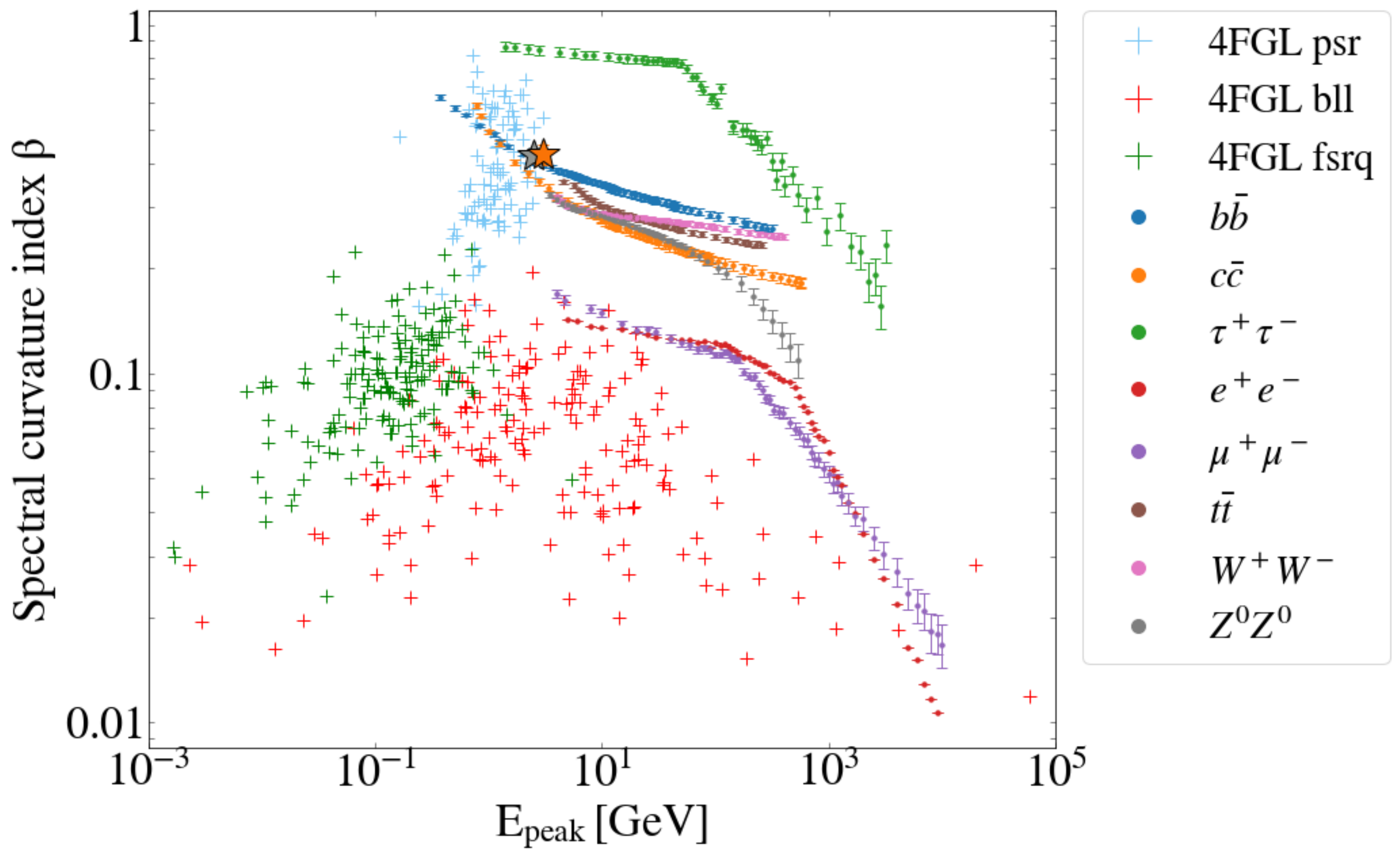}
\caption{Spectral curvature index ($\beta$) vs. the peak energy ($E_{peak}$) for 4FGL pulsars (psr, blue crosses), BL Lacs (bll, red crosses) and flat spectrum radio quasars (fsrq, green crosses) with TS of detection ($\mathrm{TS_d}$) greater than 1000. Colored data symbols with error bars correspond to the ($E_{peak}$, $\beta$) with $1\sigma$ uncertainty, for different DM annihilation channels (see legend). Large gray and orange stars correspond to the best-fit annihilation channels for the two unID sources with $\mathrm{TS_d}>1000$ that are well fitted by a LP model out of the 44 unIDs in our list.}
\label{fig:beta_plot_hights} 
\end{figure}

The test starts from the preliminary 4FGL \textit{Fermi}-LAT source catalog \cite{4fgl_paper}. More precisely, we focus on those sources in the catalog that are well fitted by a LP (see Eq. \ref{eq:logparabola}). We take the curvature spectral index, $\beta$, of the sources in the catalog and the peak energy, $E_{peak}$, i.e., the energy at which the energy flux reaches its maximum, computed from 4FGL parameters as,

\begin{equation}
\label{eq:e_peak}
E_{peak} = E_0\cdot e^{\frac{2-\alpha}{2\beta}}
\end{equation}

\noindent where $E_0$ is the pivot energy and $\alpha$ the photon spectral index. In this peculiar parameter space, for the brightest sources ($\mathrm{TS_d}>1000$\footnote{We will assume $\mathrm{TS_d}\sim\sigma^2$, where $\sigma$ is the significance of detection.}), there is a clear separation between pulsars, BL Lacs (BLLs), and flat-spectrum radio quasars (FSRQs), as shown in Figure \ref{fig:beta_plot_hights}. In particular, bright blazars present $\beta<0.2$ (i.e., less curved spectrum), while pulsars typically exhibit $0.1<\beta<1$ (higher curvature). Also, BLLs peak at higher energies than FSRQs.

The separation of classes with parameters such as the spectral curvature or the photon index has been addressed by different authors in the past \cite{Lefaucheur2017,Ackermann2012}, but none of them included DM as a possible class in their analyses. Previous work \cite{Calore+17} \cite{Coronado_Blazquez2019} showed that the PLE represents a fair approximation to the DM annihilation spectrum. Following the same arguments, and since LP exhibits a spectral curvature as well, we now perform a fit to LP for every annihilation channel. We do so, first, by producing spectral DM data points as given by PPPC4IDDM \cite{Cirelli+12}, which are then fitted with LP. The fit behaves better for hadronic channels compared to leptonic ones, as expected. We do not perform fits for $m_\chi>10$ TeV, as higher-order electroweak corrections, not included in PPPC4IDDM, may be relevant approximately above this energy. As can be seen in Figure \ref{fig:beta_plot_hights}, for each considered channel, $E_{peak}$ can look quite different  depending on the considered channel and mass. For example, for $b\bar{b}$, $E_{peak}\sim$ $m/20$ $(m/50)$ for low (high) DM masses, while $\tau^+\tau^-$ roughly peaks at $m/3$ independently of the mass. We now populate the $E_{peak}-\beta$ parameter space with the $E_{peak},\beta$ values derived from these DM fits to LP performed for different masses and channels. Interestingly, a good portion of the region of the parameter space where the DM resides is radically different from the one where astrophysical sources lie; again, see Figure \ref{fig:beta_plot_hights}.

DM and pulsars only share a region in the parameter space for DM masses ranging approximately from 5 to 70 GeV, and for the $b\bar{b}$ and $c\bar{c}$ annihilation channels. This is the region where the mentioned pulsar/DM confusion is present. We also show in Figure \ref{fig:beta_plot_hights} the result of using LP for two sources with $\mathrm{TS_d}$>1000 and $\Delta$TS>-25 (see Eq. \ref{eq:akaike}), namely 3FGL J0336.1+7500 ($Z^0Z^0$, $\mathrm{TS_d}$=1095, $\Delta$TS=-6.60) and 3FGL J1225.9+2953 ($c\bar{c}$, $\mathrm{TS_d}$=1149, $\Delta$TS=-0.02). They both lie precisely in the parameter space typical of pulsars (yet not far from the DM area).

\begin{figure}[!ht]
\centering
\includegraphics[height=9cm]{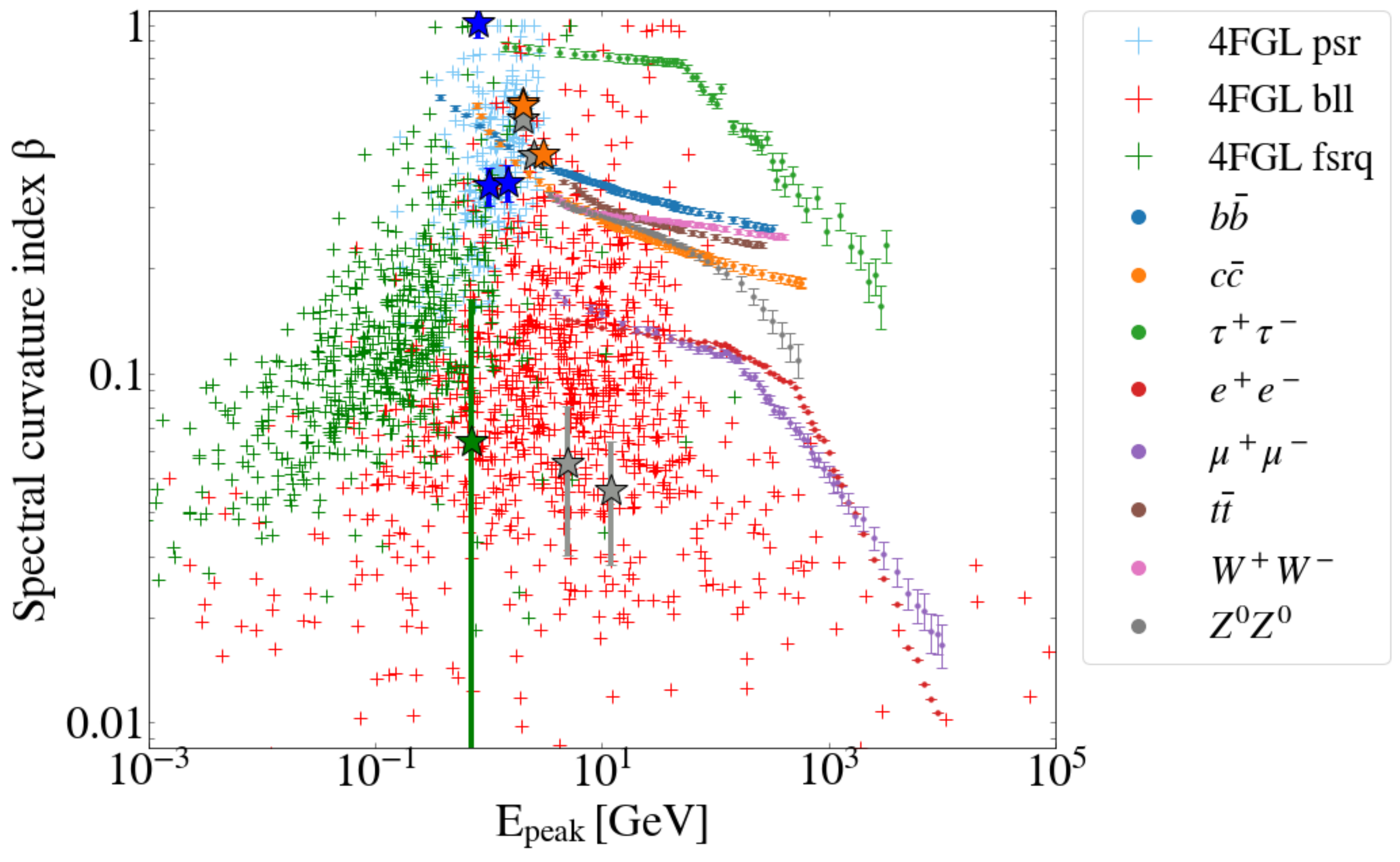}
\caption{Same as Figure \ref{fig:beta_plot_hights} but without cut in detection TS. Also shown as stars are those unIDs that are well fitted by LP (i.e., $\Delta\mathrm{TS}\geq-25$). The color of each star refers to the best-fit DM channel in each case; see legend.}
\label{fig:beta_plot_allts}
\end{figure}

The test described above is particularly powerful for bright sources. In Figure \ref{fig:beta_plot_allts}, we repeat the same test for all sources instead, i.e., without a cut in $\mathrm{TS_d}$. As can be seen, the same parameter space gets very crowded when including weak sources. For instance, now the blazars' and quasars' curvature spectral indices range up to 1, though in all these cases $E_{peak}$ still remains below 100 GeV with a few exceptions (and, interestingly, all of them with $\beta<0.1$). Also, the pulsars keep confined to $\beta>0.1$, with the exception of only two sources which lie very close to this value. All in all, for weak sources the $E_{peak}-\beta$ parameter space can still be useful to shed light on the type of objects, e.g. by rejecting the pulsar interpretation for objects exhibiting $\beta<0.1$ or, alternatively, $E_{peak}>10$ GeV.

From a DM perspective, the proposed test will be particularly valid for energies higher than 100 GeV, i.e., where no astrophysical source peaks but DM can. This roughly corresponds to DM masses larger than $\sim$1 TeV (depending on the specific annihilation channel). Figure \ref{fig:beta_plot_allts} also plots the 11 sources with a fit to DM ($\Delta\mathrm{TS}\geq-25$) which are also well fitted with LP. None of these sources lie in the region of the parameter space exclusive to DM. Yet, it is possible to draw some useful conclusions from the DM perspective from this exercise. Three unID sources (3FGL J0539.2$-$0536, 3FHL J0343.5$-$6302 and 3FHL J1440.2$-$2343) exhibit values of $\beta$ much lower than the ones expected for DM annihilation, i.e., $\sim$0.05 versus values around $0.3-0.9$, respectively. Also, the 1-$\sigma$ error bars, with $\mathrm{\Delta}\beta$ ranging from $\sim$0.025 to 0.1, are not large enough to reach the DM parameter space. These sources may be discarded as DM subhalos, although as the statistical significance of these uncertainties is unclear, we conservatively do not discard any of them for this reason. Although two of them are already discarded for not fitting any DM channel with $\Delta$TS>-9, it is worth remembering that 3FHL J0343.5$-$6302 is one of the sources with $\Delta$TS(DM)>0 (see Table \ref{tab:sources_summary}). Although the uncertainty in $\beta$ is high, all these unIDs are well below the expected DM $E_{peak}-\beta$ values. Also, we predict these sources to be most likely blazars.

With respect to the remaining seven unID sources, well fitted by $b\bar{b}$, $c\bar{c}$ and $Z^0Z^0$, they all lie close to the DM region, thus conservatively we cannot discard a DM hypothesis or classify them either as pulsars or blazars due to the high confusion in this part of the parameter space for the case of weak sources.

As a summary, this new discriminating tool based on the $E_{peak}-\beta$ parameter space can potentially discard unIDs as DM subhalos, when these exhibit low ($E_{peak}$, $\beta$) values of order ($0.01-0.3, 0.01-1$), or to point out particularly interesting DM  candidates. Indeed, there is a specific region of the parameter space that is only reachable by DM annihilation with high ($E_{peak}$, $\beta$), roughly larger than ($0.1-1$,$1-10$). No unIDs among our list of best candidates are found there though. A more quantitative assessment of the boundaries of each source class in this parameter space is under study using machine learning techniques and will be published elsewhere.

\section{Spatial analysis of remaining DM candidates}
\label{sec:spatial_analysis}

\subsection{Technical setup and data analysis results}
\label{sec:tech_spatial}
Spatial extension has been studied by many authors as a possible ``smoking gun'' for DM annihilation (see e.g. \cite{Bertoni+15,Bertoni+16,ZechlinHorns12,Huetten2018,Chou2017,FHES_paper}). In this section we search for spatial extension, in the 7 best DM candidates ($\Delta\mathrm{TS}>0$ in Eq. \ref{eq:akaike}) that are left after the spectral analysis in \cref{sec:spectral_section}. The spatial extension is defined as the 68\% containment width of the signal. We choose two different templates, a 2D uniform disk and a Gaussian profile. The {\tt GTAnalysis.extension} \textit{fermipy} module fits the source with the mentioned spatial profiles in concentric circles of increasing radii, centered at the position of the source, and computes the likelihood values for each of them. The fit is computed in 40 linear radial bins between 0.01 and $0.5^\circ$.

We perform a TS analysis to study the preference of the extended model over a point-like source, similarly to what done via Eq. \ref{eq:TS} to estimate detection significances, but where now $H_1$ represents the extended source model (alternative hypothesis) and $H_0$ the point-like source (null hypothesis). The latter is defined as the one with an extension smaller than the point spread function (PSF) at the energy under consideration. On the other extreme, should significant extension be found, larger than the maximum allowed angular size of $0.5^\circ$ (i.e. the log-likelihood is monotonically increasing in that range), the latter upper limit is increased. 

\begin{figure}[!ht]
\centering
\includegraphics[height=4.48cm]{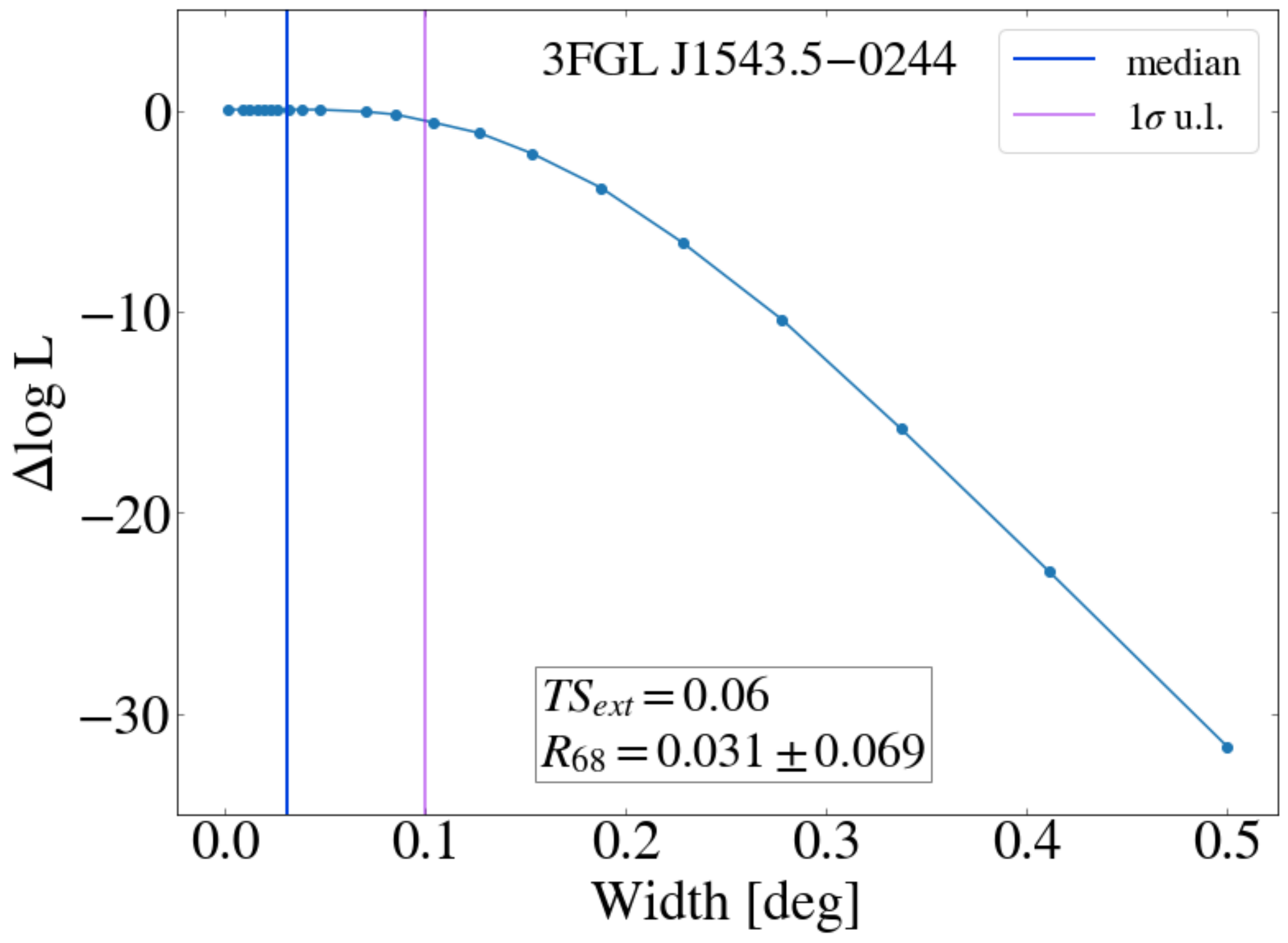}
\includegraphics[height=4.48cm]{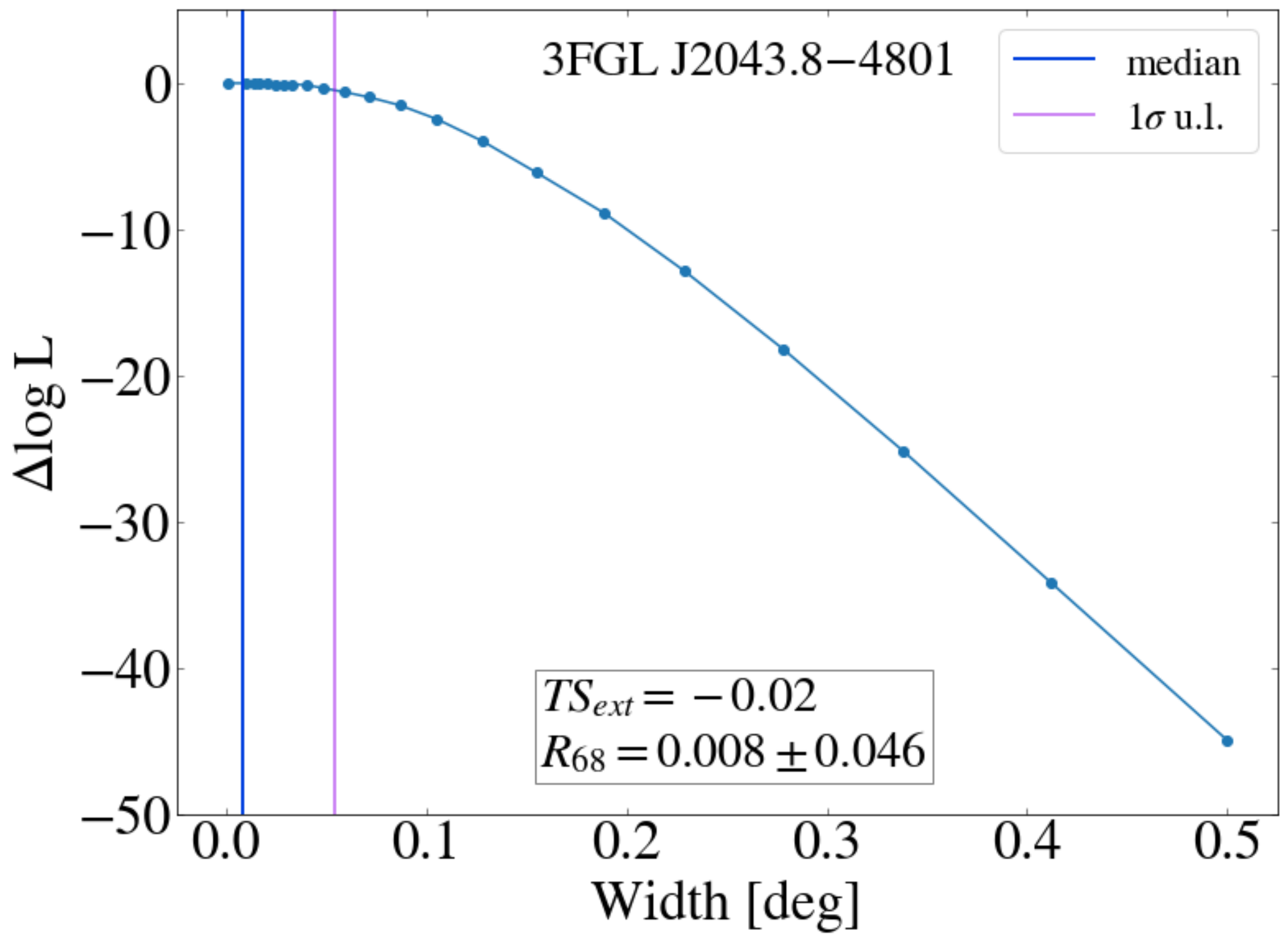}
\vfill
\includegraphics[height=4.48cm]{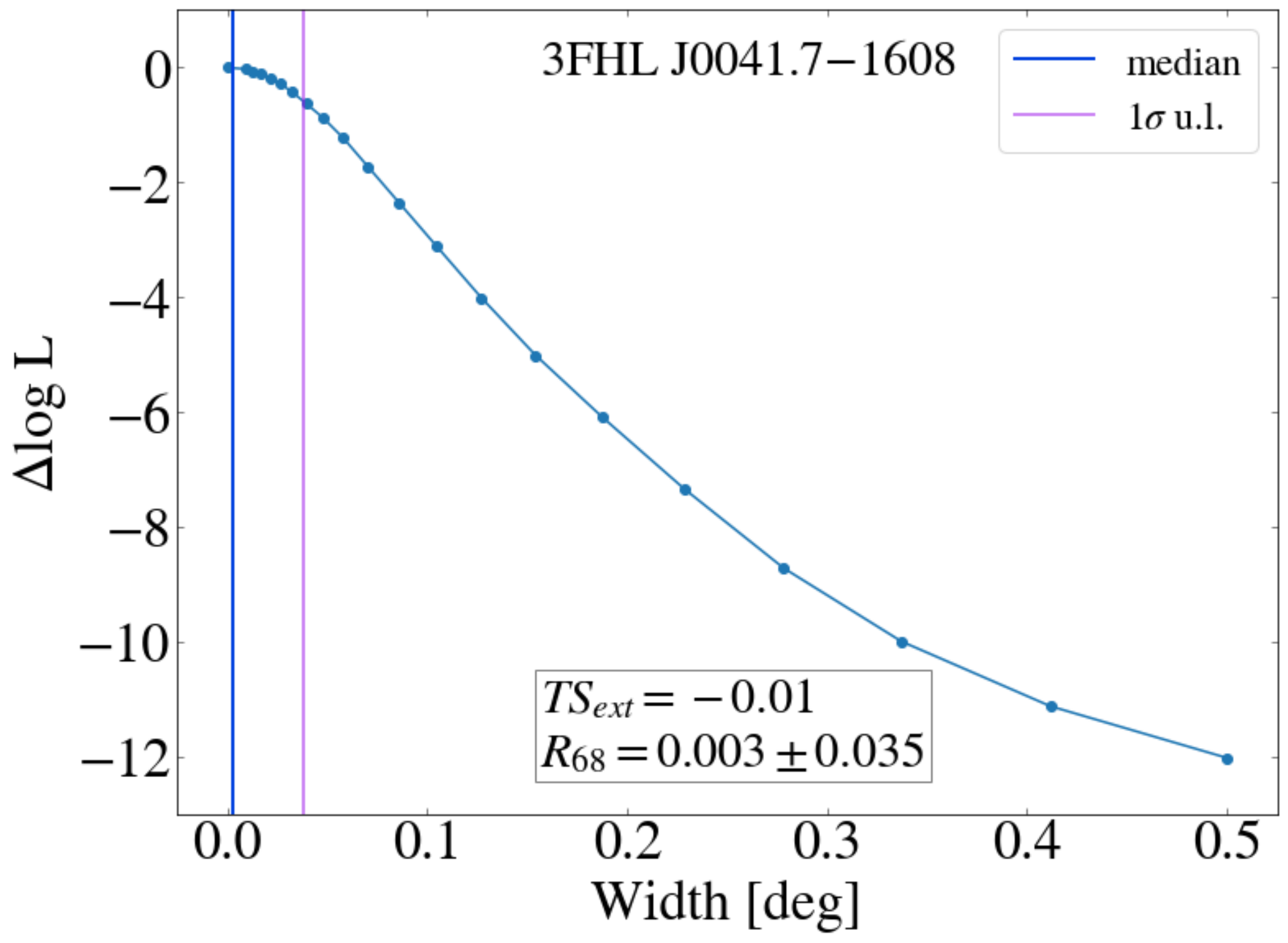}
\includegraphics[height=4.48cm]{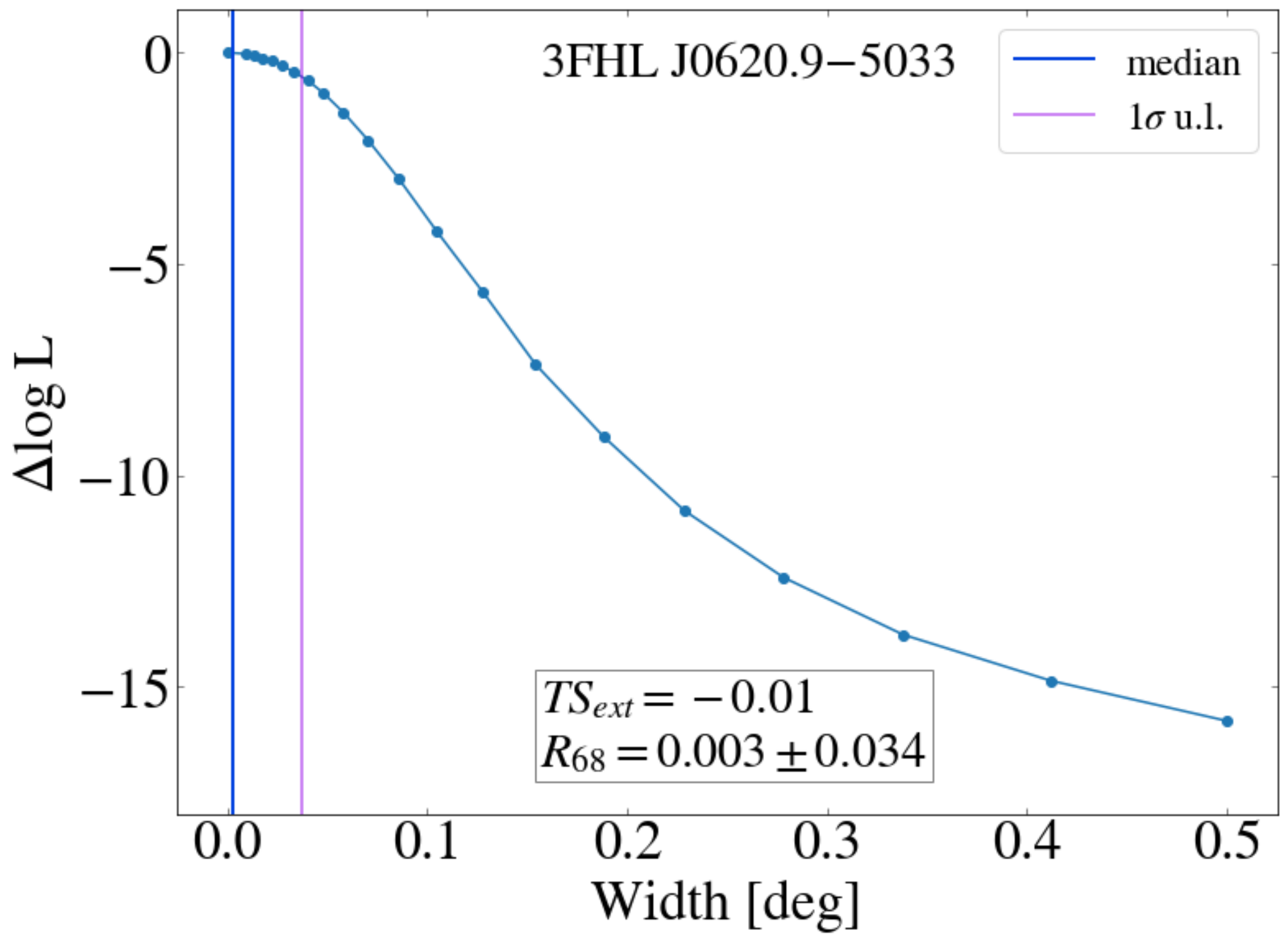}
\vfill
\includegraphics[height=4.48cm]{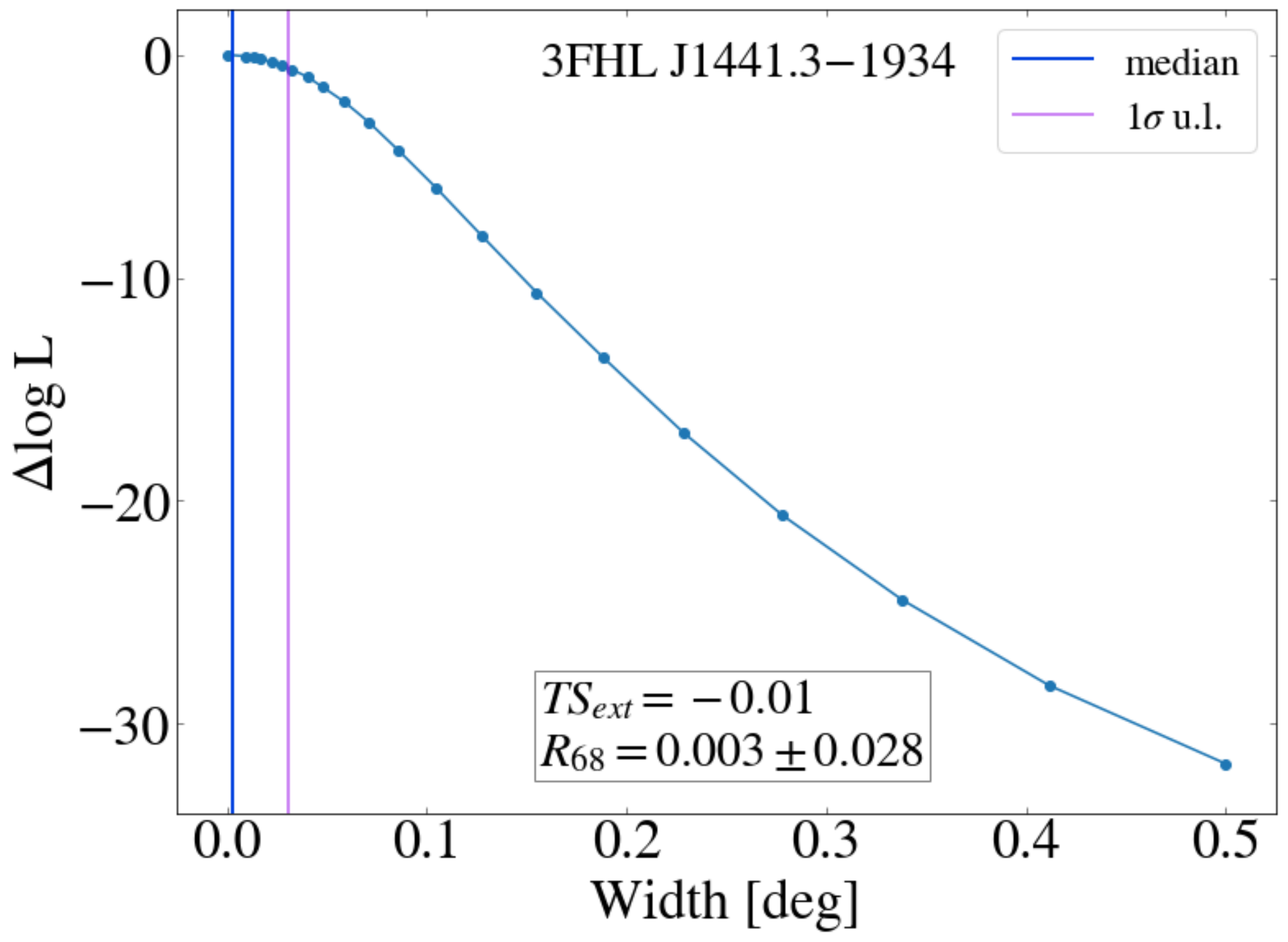}
\includegraphics[height=4.48cm]{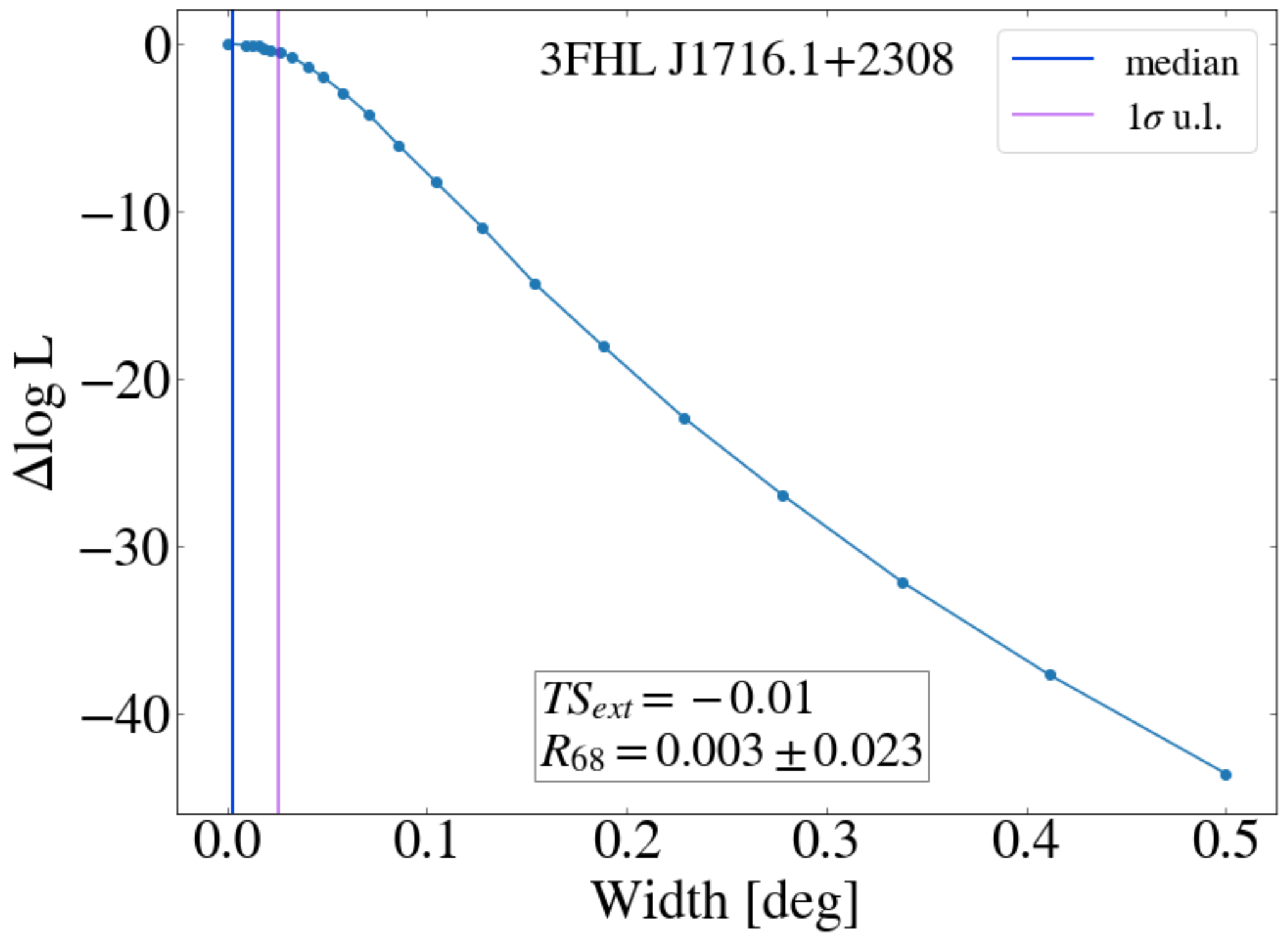}
\vfill
\includegraphics[height=4.48cm]{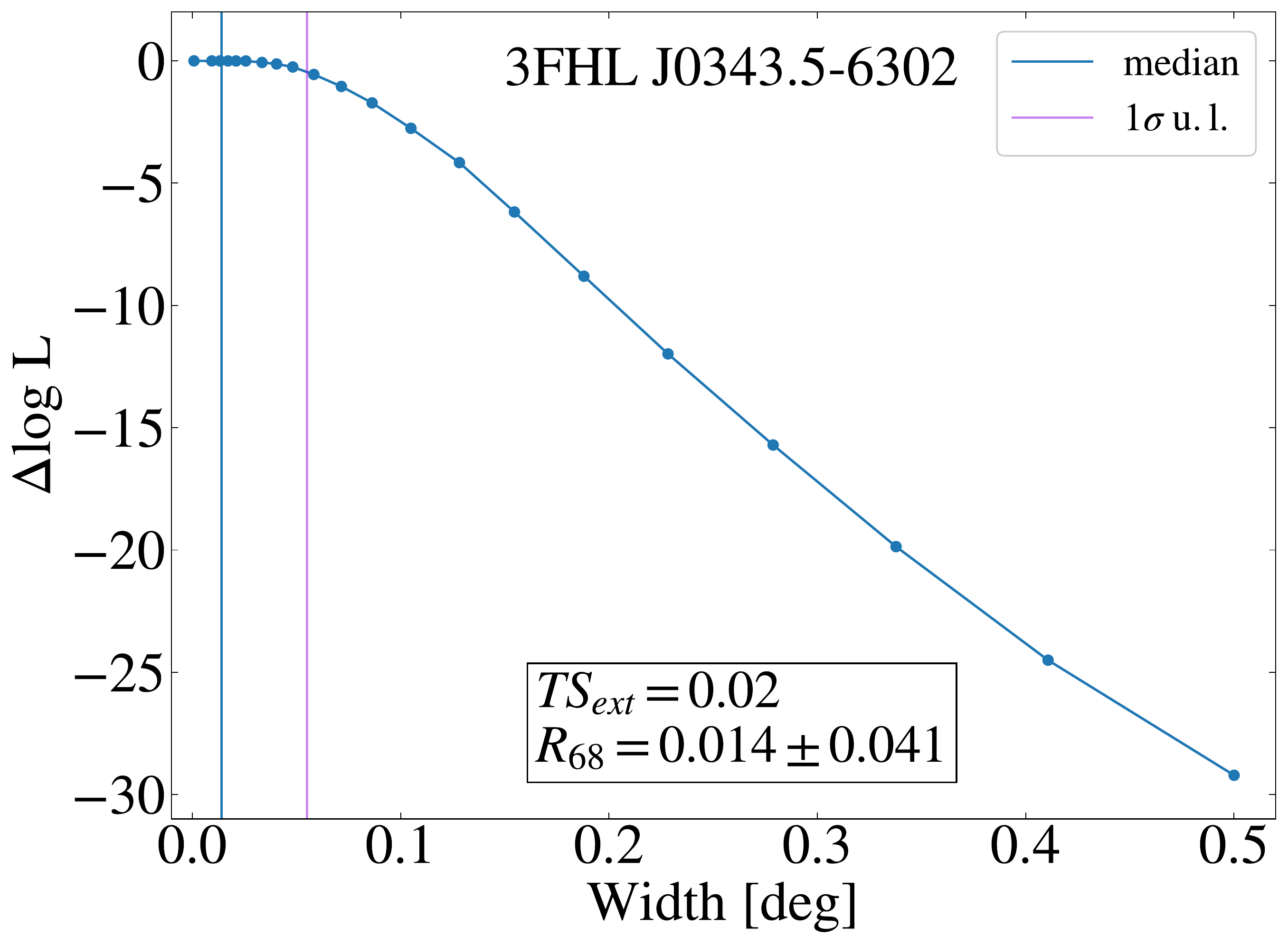}
\includegraphics[height=4.48cm]{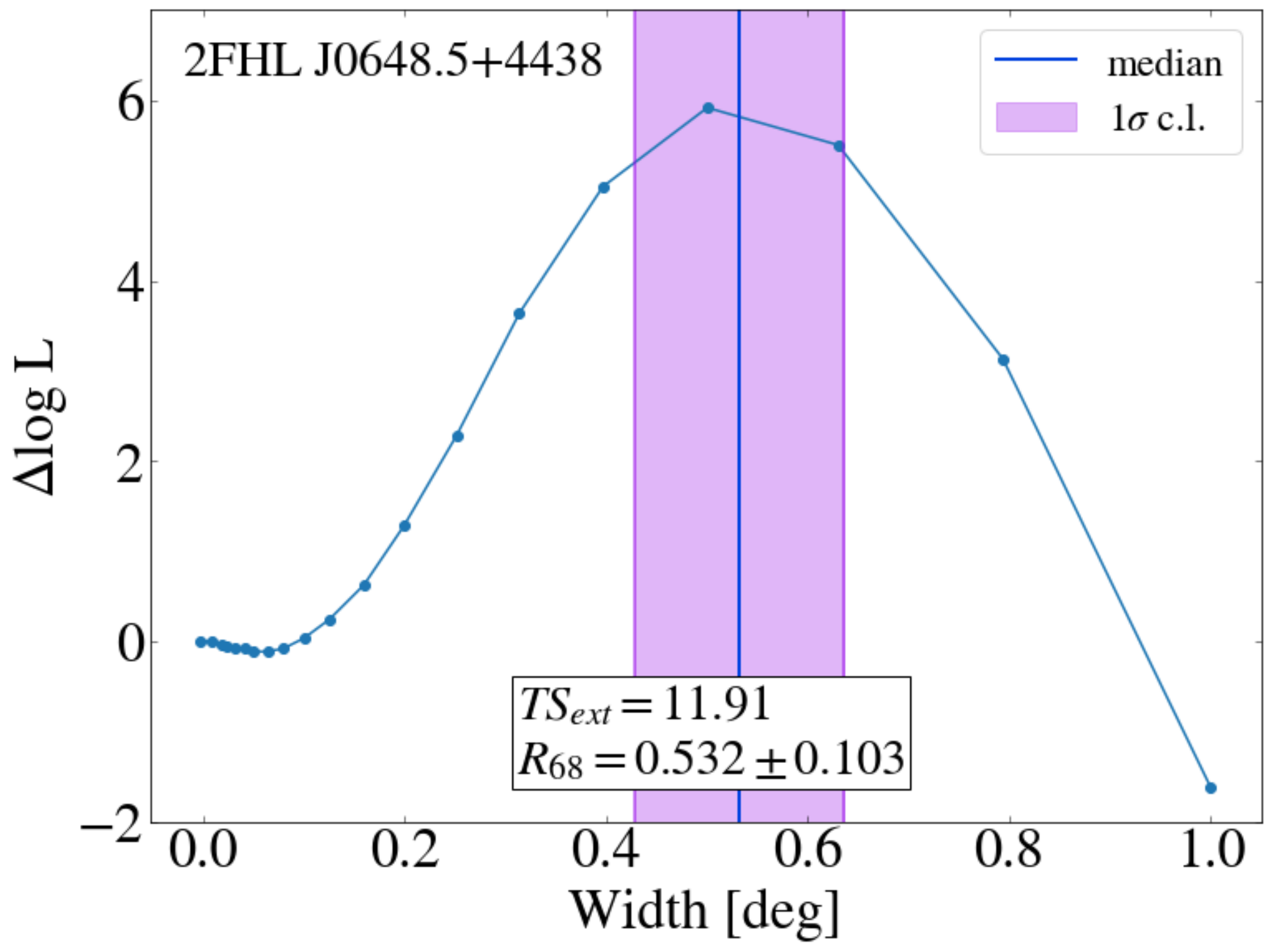}
\caption{Spatial extension log-likelihood profiles of the seven DM candidates from Table \ref{tab:sources_summary} and one known extended source, for the case of adopting a Gaussian spatial template. The blue, vertical line is the median best-fit value, while the violet line refers to the $+1\sigma$ uncertainty of the mean (in the last source, where there is a preference for extension, this is an uncertainty band). All DM candidates appear as point-like. The use of a 2D disk spatial template instead of a Gaussian one leads to similar results. Only for illustrative purposes, we also show in the bottom right panel the source 2FHL J0648.5+4438, which is not well fitted to any DM channel but is found to have a 68\% containment width of $0.5\pm0.1^\circ$ at $\sim$3$\sigma$ significance.}
\label{fig:ext_likelihood}
\end{figure}

None of the candidates is found to have significant spatial extension. Figure \ref{fig:ext_likelihood} shows the log-likelihood profiles and corresponding TS values of the extension evaluated with the Gaussian profile. Changing the spatial template to the 2D disk provides similar results.

\subsection{$\Lambda$CDM predictions of subhalo angular sizes}
\label{sec:interpretation}

State-of-the-art N-body simulations of Milky Way (MW) size halos have become a powerful tool to study halo substructure and have provided the most valuable and accurate information on the properties of the present-day subhalo population \cite{Springel+08,vlii_paper,GHALO,Elvis}. In \cite{Coronado_Blazquez2019}, we used publicly available results from the Via Lactea II (VL-II) DM-only simulation at redshift zero\footnote{\url{http://www.ucolick.org/\~diemand/vl/}} to perform this task. More specifically, we obtained $\Lambda$CDM predicted annihilation fluxes of the Galactic subhalos to, ultimately,  set DM constraints by comparing predictions with data. To do so, and as described in full detail in \cite{aguirre2019}, we first replicated the original VL-II and repopulated it with millions of low-mass subhalos below the resolution of the parent simulation. We then placed the observer at arbitrary positions at 8.5 kpc from the Galactic center in order to obtain multiple realizations. 

\begin{figure}[!ht]
\centering
\includegraphics[height=11cm]{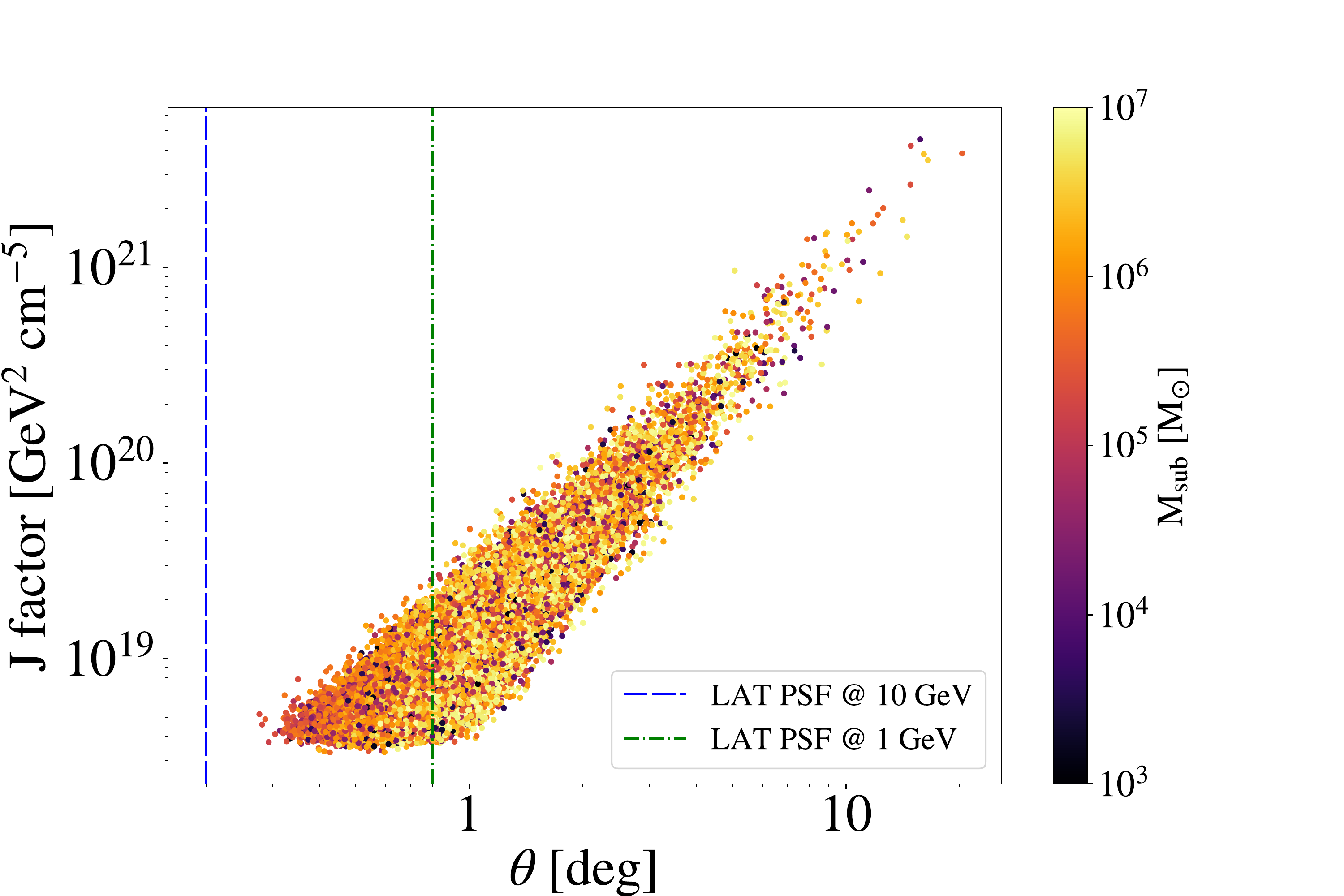}
\caption{Angular extension of subhalos following our definition in Eq.~\ref{eq:ang_extension} as predicted by the N-body simulation work of \cite{aguirre2019} (see also \cite{Coronado_Blazquez2019}). The figure shows the 100 brightest subhalos in each of the 1000 realizations of the repopulated VL-II simulation,  which include subhalos with masses down to $10^3 M_{\odot}$, i.e. well below the resolution of the original VL-II. The typical \textit{Fermi}-LAT PSF at $E>1$ GeV and $E>10$ GeV are also shown as vertical dashed lines. Most subhalos are expected to be extended for \textit{Fermi}-LAT, especially the brightest ones.}
\label{fig:ang_extension}
\end{figure}

Yet, in \cite{Coronado_Blazquez2019} we did not study the \textit{typical} angular extent of subhalos in the sky. This feature may turn out to be relevant for the current work, for which the spatial extension of the best DM subhalo candidates among our list of remaining unIDs was analyzed. Thus, we now compute the angular size of subhalos in our realizations of the repopulated VL-II as well. We do so by defining it in the following way:

\begin{equation}
\theta_{sub} [\text{deg}] = \frac{180}{\pi} \text{atan} \left( \frac{r_s(m_{sub})}{ D_{Earth}} \right)
\label{eq:ang_extension}
\end{equation}

\noindent where $r_s$ is the scale radius of the subhalo, with mass $m_{sub}$, when a Navarro-Frenk-White (NFW) DM density profile is assumed \cite{1997ApJ...490..493N}, and $D_{Earth}$ is the distance to the Earth. We recall that for an NFW profile 90\% of the total annihilation flux originates within $r_s$. As subhalos exhibit a truncated NFW (see, e.g., Ref.~\cite{Kazantzidis2004,Moline+17}), the use of $r_s$ in Eq.~\ref{eq:ang_extension} represents a very good approximation for the size of the region where most of the DM flux was originated. The angular extension for the 100 brightest subhalos below $10^7~M_{\odot}$ in 1000 repopulated VL-II realizations is shown in Figure \ref{fig:ang_extension}. The typical \textit{Fermi}-LAT PSF at $E > 10$ GeV, $\sim0.2º$, is also shown in the same figure\footnote{\url{http://www.slac.stanford.edu/exp/glast/groups/canda/lat_Performance.htm}}. According to this figure, every subhalo should appear as extended for the LAT. However, no unID was found to be extended in our spatial analysis of \cref{sec:tech_spatial}. In principle, this means that none of the seven best candidates are DM subhalos. Yet, as these unIDs correspond to faint sources, it is not totally clear that they would appear as extended in our analyses even if being actual DM subhalos. We conclude that more work is needed before being able to use $\Lambda$CDM predictions of subhalo angular sizes to rule out unIDs as potential DM subhalos. This work will be done elsewhere. For the time being, in the spirit of the conservative approach of this work, we do not reject any source as DM subhalo for this reason.

The brightest subhalos in Figure~\ref{fig:ang_extension} are all expected to be significantly extended, with typical sizes of the order of $10^\circ$. Also, some of them are expected to be light, with masses as low as $10^5 M_\odot$ or even less. Thus, these small subhalos should be close to Earth ($\sim2$ kpc) in order to be among the brightest ones. These subhalos, with the largest predicted angular sizes of $\mathcal{O}(10^\circ)$, are the ones used to set the DM constraints in \cref{sec:constraints}.

\section{Search for Stellar Counterparts with \Gaia\ DR2}
\label{sec:Gaia_analysis}
Our DM candidate sources in Table~\ref{tab:sources_summary} have no obvious associated stellar counterparts such as dwarf galaxies. If a dwarf galaxy is observed in the same region as a gamma-ray source which has a similar spectrum to what is expected from the DM annihilation, it would provide stronger support for the gamma-ray source to be a ``smoking gun'' for DM annihilation \citep[e.g.][]{Bertoni+15,Bertoni+16,Calore+17}. The European Space Agency's \Gaia\ mission \citep{Gaia+Prusti+16} has made the second data release of their unprecedented parallax and proper motion measurements for more than one billion stars brighter than $G<20$ mag \citep[\Gaia\ DR2;][]{Gaia+Brown+18,Lindegren+18}. This provides us with a new window to find an overdensity of stars, which could be dwarf galaxies, using proper motions in the inner Galactic halo, where the stellar density is too high to detect them from the photometric data alone \cite{Antoja+15}. In \cite{Ciuca18} the authors used both parallax and proper motion data from \Gaia\ DR2, and searched for a dwarf galaxy in the fields of the eight \textit{Fermi}-LAT extended sources suggested by \cite{Bertoni+15,Bertoni+16,Xia+17,FHES_paper}. Unfortunately, they did not find any stellar counterpart in the \Gaia\ data. Still, using mock data, \cite{Ciuca18} obtained a conservative limit on the stellar mass of $M_* < 10^4$~M$_{\odot}$ for $d < 20$\, kpc, which provided a constraint on the potential stellar counterpart. However, their targeted sources are either not found in the follow-up studies \cite{FHES_paper} or are located at lower Galactic latitudes. Hence, it is interesting to apply this new window of searching dwarf galaxies to our DM candidate sources that are carefully selected from the improved spectral analysis.

Following the approach of \cite{Ciuca18}, we selected a region of the sky within 2$º$ of each of the best seven dark matter candidates shown in Table \ref{tab:sources_summary}, namely 3FGL J1543.5$-$0244 ($l = 4.07^\circ$, $b = 38.90^\circ$), 3FGL J2043.8$-$4801 ($l = 351.75^\circ$, $b = -38.47^\circ$),  3FHL J0041.7$-$1608 ($l=110.90^\circ$, $b=-78.79^\circ$), 3FHL J0343.5$-$6302 ($l=277.20^\circ$, $b=-44.54^\circ$), 3FHL J0620.9$-$5033 ($l=258.80^\circ$, $b=-25.29^\circ$), 3FHL J1441.3$-$1934 ($l=335.55^\circ$, $b=36.24^\circ$) and 3FHL J1716.6+2308 ($l=45.04^\circ$, $b=30.64^\circ$), from the data in \Gaia\ DR2. We then used both spatial and proper motion information from \Gaia\ DR2 to find the stellar signature of a co-moving stellar component at the position of each field of our DM candidate sources. Depending on the distance at which we are searching for a stellar counterpart, i.e. either closer or farther than $d=10$\ kpc, we apply a parallax filter to minimize the background contamination. In the case of search distances greater than $10$\ kpc, we apply the same parallax filter used in \cite{Antoja+15}, where we discard stars for which $\varpi - \sigma_{\rm \varpi} > 0.1$\,mas, where $\varpi$ represents the parallax and $\sigma_{\rm \varpi}$ is the parallax uncertainty. This filtering is equivalent to removing stars closer than 10\,kpc within parallax uncertainties and aims to minimize contamination from foreground stars. If the dSph is assumed to be closer than $d < 10$\,kpc, we employ a distance-dependent parallax cut, namely $1/(2 d_{\rm in})< \varpi < 1.0/d_{\rm in}$, which changes as we are probing a distance range between 1.0 and 10\,kpc in increments of $d_{\rm in}$ of 1\,kpc up to 5 kpc, and then 2~kpc up to 9 kpc.

Following \cite{Ciuca18}, we used an Extreme-Deconvolution \citep[XD,][]{Bovy+11} Gaussian Mixture Model \citep[XDGMM,][]{Holoien+17}, to detect a stellar overdensity associated with a stellar counterpart as a density enhancement in the four-dimensional data vector comprised of the stellar position in Galactic longitude and latitude, and the RA and DEC proper motion measurements as we are looking to find a concentration of stars both in proper motion space and in the sky. \cite{Ciuca18} demonstrated that this method allows us to take into account the correlation between the measurement of RA and DEC proper motions. As done in \cite{Ciuca18}, for practical purposes, we applied a small constant uncertainty of $0.01^\circ$ in the stellar positions, whose effect on our results is thought to be insignificant, and we do not take into account the correlations between proper motion and position as they are very small.

\begin{figure}
	\centering
	\includegraphics[width=\hsize]{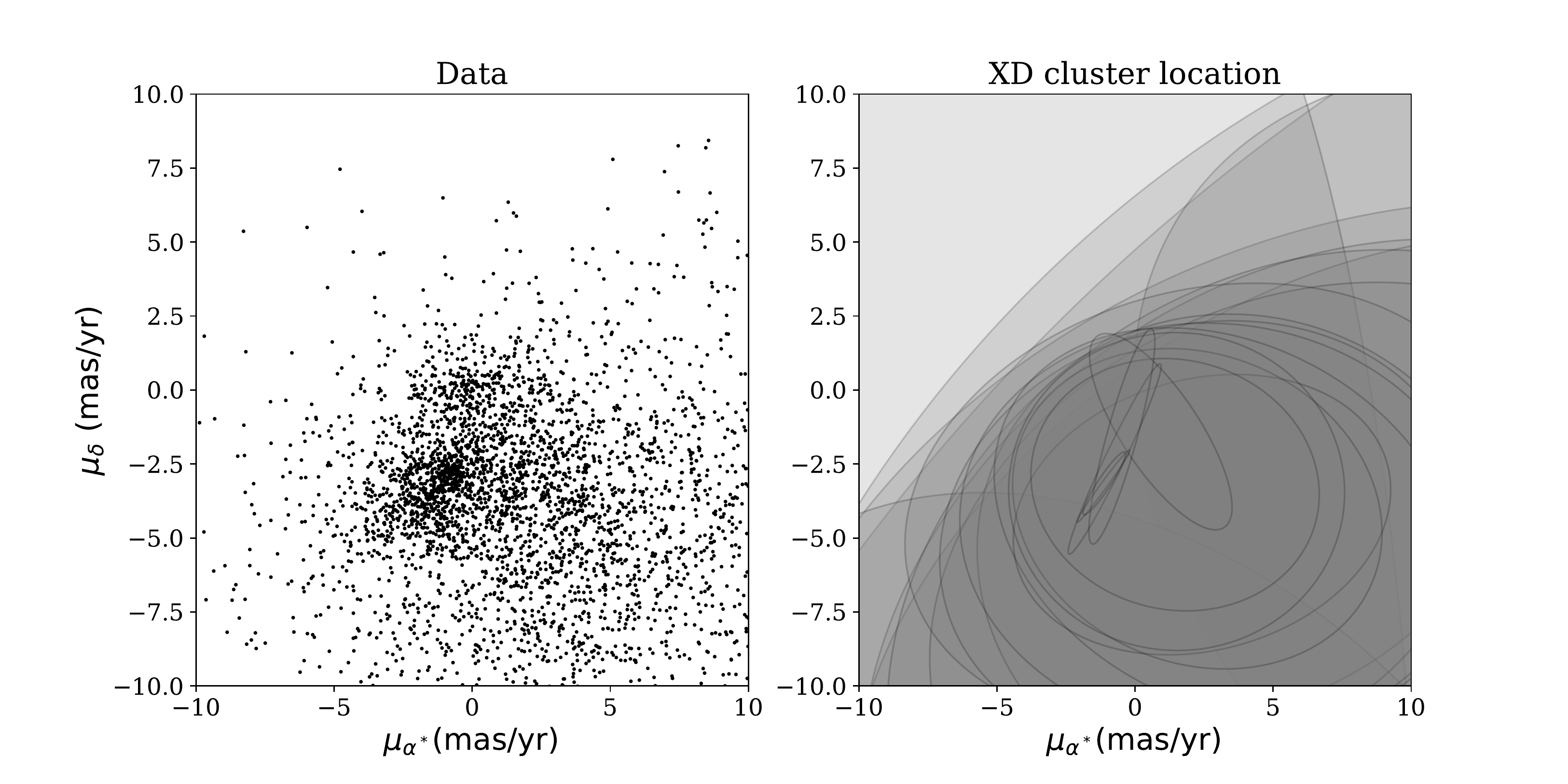}
	\caption{A detection of a kinematical structure in the field of 3FHL J0041.7$-$1608. The left panel shows the data after filtering with $d>10$~kpc. The right panel shows the results from the Gaussian Mixture Model \citep{Holoien+17} used to perform density estimation using the XD method \citep{Bovy+11} with 25 components. Ellipses represent detected Gaussian components to describe the data and the darker gray color indicates more components overlapping. The small ellipse around $(\mu_{\alpha^*}, \mu_{\delta})=(-1.12,-3.15)$~mas yr$^{-1}$ is believed to be part of the Sagittarius stream in the right panel.
} 
	\label{fig:sagcandidate}
\end{figure}

We used an XDGMM with 25 four-dimensional Gaussian components, to retrieve the structure present in the data, to the samples of stars with d < 10 kpc and d > 10 kpc, filtered as mentioned above, in the seven fields of our DM candidate sources. 25 Gaussian components are found to be a large enough number of components to detect a co-moving stellar structure in the relatively high Galactic latitude regions like our target regions \cite{Ciuca18}. In that same work, the detectability of a dwarf galaxy depending on the number of Gaussian components by placing a synthetic mock dwarf galaxy data in the background data created using the Gaia data with the same size as that of the field of view used for our search is tested, and demonstrated that the detected mock dwarf galaxy showed up as a narrow Gaussian component in the RA and DEC proper motions. In fact, an excessive number of components in the XDGMM does not penalize the detection of the dwarf galaxy (see also \cite{Ciuca18,Anderson2017}). Therefore it is not necessary to search for the optimum number of the components, rather using a large enough number of the Gaussian components is sufficient.

We found no evidence for any co-moving stellar group in the fields, except in the field of 3FHL J0041.7$-$1608. A clear overdensity of stars highlighted by a small narrow ellipse is detected in the proper motion space as shown in Figure \ref{fig:sagcandidate}. However, we believe that this is not evidence of a dwarf galaxy associated with 3FHL J0041.7$-$1608. It is worth noticing that this is a part of the known Sagittarius stream. Our detected proper motion is $(\mu_{\alpha^*}, \mu_{\delta})=(-1.12,-3.15)$~mas~yr$^{-1}$, which is consistent with the expected proper motion from the Sagittarius stream (e.g. see \cite{Hayes2018}). We also explored the proper motion distribution in a much larger region of the sky around this source (as far as 90$^{\circ}$ away), and found the excess of the proper motion to be extended in the large region of the sky where the Sagittarius stream is expected in the Southern Galactic region (e.g. see \citep{Mateu+2018MNRAS.474.4112M} for a compilation of the known stellar streams). It is interesting to find that one of our DM candidate sources coincides with the Sagittarius stream. However, considering the fraction of the sky covered by the Sagittarius stream, this coincidence is not surprising. Still, this remains a compelling fact, although it is difficult to confirm or reject that this is an annihilation emission signal from a DM subhalo in the Sagittarius stream.

\section{Conclusions and prospects}
\label{sec:conclusions}

In this work, we have presented a dedicated LAT analysis of those unIDs that were identified as the best DM subhalo candidates within the latest \textit{Fermi}-LAT catalogs by \cite{Coronado_Blazquez2019}. Now, we are able to discard most of these candidates, mainly based upon their spectral information.

First, we performed a full spectral analysis with \textit{Fermipy}, using all 10-year available gamma-ray data. With the new data set, not all our previous candidates reach the standard detection threshold of $\mathrm{TS_d}=25$. In fact, by just improving the LAT statistics from 48 to 116 months, and by using the new multi-wavelength information available, we are able to reduce the number of candidates in 9 and 3 sources, respectively. More precisely, we are left with 32 unIDs out of an initial poll of 44. For these remaining 32 unIDs, we performed a spectral analysis in the 300 MeV-800 GeV range. Three different parametric models were used to perform the SED fitting: a power law, log-parabola and a power law with super-exponential cutoff. Also, we used \texttt{DMFit} to fit the SED to different DM annihilation channels. The goal of the fitting procedure was to find the spectral model that best fits the data and, thus, to understand if a DM origin was statistically preferred over typical astrophysical gamma-ray emitters for any of the studied unIDs.

We found a significant number of sources that preferred a non-exotic, astrophysical origin at high ($>3\sigma$) significance. These sources were rejected from our list of potential DM subhalos. On the other hand, only seven unIDs (two 3FGL, five 3FHL and zero 2FHL sources) were found to marginally prefer a DM origin for at least one annihilation channel. Yet, in all cases the slight preference for DM was found to not be statistically significant. 

After our spectral analysis, that allows us to disregard some of the unIDs as due to DM annihilation, and in the absence of outstanding DM subhalo candidates in the list of remaining sources, we set stringent constraints on the $m_{\chi}$-$\langle\sigma v\rangle$ parameter space. We do so by adopting the same methodology that was presented in \cite{Coronado_Blazquez2019}, which is based on a comparison between number of predicted observable subhalos from simulations, and actual number of remaining unIDs in our catalogs. By adopting a conservative view, we keep in our list of potential DM subhalo candidates all those unIDs that do {\it not} exhibit a preference for an astrophysical spectral model at more than $3\sigma$. Our derived DM constraints rule out canonical thermal WIMPs up to $\sim$10 GeV for $b\bar{b}$ and $\sim$20 GeV for $\tau^+\tau^-$ annihilation channels.

In \cref{sec:beta}, we described a new test that can be potentially used for discriminating astrophysical sources from DM. This can be particularly interesting in the case of disentangling low-mass WIMPs from pulsars. The method is based on the parameter that controls the spectral curvature in the log-parabola SED model, $\beta$, and on the peak energy of the source, $E_{peak}$. Sources of a particular type tend to cluster in this particular ($E_{peak},\beta$) parameter space, indeed astrophysical sources and DM populating different regions of it. Although the discrimination power of this ``$\beta$ plot'' behaves differently depending on the TS of detection (the larger its value the better for discrimination), we are able to mark one of our best candidates as very unlikely due to DM annihilation, although we conservatively do not reject it due to the uncertain statistical significance. In the rest of cases the test is not conclusive. Work is already ongoing to further scrutinize the nature of unIDs using this promising new tool for source type discrimination. 

In \cref{sec:spatial_analysis} we searched for spatial extension of our seven best candidates, using two different spatial templates (an uniform 2D disk and a Gaussian profile). All of the candidates are found as point-like sources. Interestingly, our N-body simulation work predicts that the brightest members of the subhalo population should exhibit a large extension, of the order of several degrees. It is unclear though that we can safely use these results to reject unIDs as potential subhalos so we decide not to do so and stay conservative: weak unIDs may appear as point sources even if they were intrinsically extended. Further work will be presented to properly address this issue as well.

Finally, using $Gaia$ DR2 and an advanced density estimation technique, we search for kinematical signature of a stellar counterpart in the fields of our seven best DM candidates listed in Table \ref{tab:sources_summary}. We find that one of the fields, the one surrounding 3FHL J0041.7$-$1608, coincides with the Sagittarius stream. This may point to a potential association of the DM annihilation signal with a hidden DM subhalo within the tidal stream of the Sagittarius dwarf galaxy. However, considering the large sky coverage of the Sagittarius stream, this is likely to be just a coincidence.

Looking into the imminent future, the release of new \textit{Fermi}-LAT catalogs (e.g. 4FGL \cite{4fgl_paper}) will provide new targets to work with and better photon statistics. The 4FGL will allow for a better discrimination power in the $\beta$-plot that could be further used to remove additional sources. Current \textit{Imaging Atmospheric Cherenkov Telescopes} (IACTs) and, very especially, the future Cherenkov Telescope Array (CTA) \cite{CTA_science_paper} could also be used to perform dedicated analyses of the best DM subhalo candidates at higher energies. In this sense, also ongoing or future observational campaigns at other wavelengths, such as X-ray or radio, may shine a light on the origin of some of the candidates presented in this work. Moreover, instruments with an improved angular resolution, such as AMEGO \cite{AMEGO_paper}, or even CTA itself, may further explore the potential spatial extension of the unIDs.

\acknowledgments

The authors would like to thank Jean Ballet and Jan Conrad for their valuable help on the statistical aspects of this work.

JCB and MASC are supported by the {\it Atracci\'on de Talento} contract no. 2016-T1/TIC-1542 granted by the Comunidad de Madrid in Spain. AAS is very grateful to the IFT Centro de Excelencia ``Severo Ochoa'' Spanish program under reference SEV-2016-0597. MDM acknowledges support by the NASA {\it Fermi} Guest Investigator Program 2014 through the {\it Fermi} multi-year Large Program N. 81303 (P.I. E.~Charles) and by the NASA {\it Fermi} Guest Investigator Program 2016 through the {\it Fermi} one-year Program N. 91245 (P.I. M.~Di Mauro). IC and DK acknowledge the support of the UK’s Science and Technology Facilities Council (ST/N000811/1 and ST/N504488/1). AD thanks the support of the Ram{\'o}n y Cajal program from the Spanish MINECO. The work of JCB, MASC and AAS was additionally supported by the Spanish Agencia Estatal de Investigación through the grants PGC2018-095161-B-I00, IFT Centro de Excelencia Severo Ochoa SEV-2016-0597, and Red Consolider MultiDark FPA2017-90566-REDC.

The Fermi LAT Collaboration acknowledges generous ongoing support from a number of agencies and institutes that have supported both the development and the operation of the LAT as well as scientific data analysis. These include the National Aeronautics and Space Administration and the Department of Energy in the United States, the Commissariat `a l’Energie Atomique and the Centre National de la Recherche Scientifique / Institut National de Physique Nucl\'eaire et de Physique des Particules in France, the Agenzia Spaziale Italiana and the Istituto Nazionale di Fisica Nucleare in Italy, the Ministry of Education, Culture, Sports, Science and Technology (MEXT), High Energy Accelerator Research Organization (KEK) and Japan Aerospace Exploration Agency (JAXA) in Japan, and the K. A. Wallenberg Foundation, the Swedish Research Council and the Swedish National Space Board in Sweden. Additional support for science analysis during the operations phase is gratefully acknowledged from the Istituto Nazionale di Astrofisica in Italy and the Centre National d'Etudes Spatiales in France. This work performed in part under DOE Contract DE- AC02-76SF00515

This research made use of Python, along with community-developed or maintained software packages, including IPython \cite{Ipython_paper}, Matplotlib \cite{Matplotlib_paper}, NumPy \cite{Numpy_paper} and SciPy \cite{scipy_paper}. This work made use of NASA’s Astrophysics Data System for bibliographic information.

\appendix

\section{Rejection criteria for the unIDs}
\label{app:rejection}
In Table \ref{tab:rejection_table} we detail the rejection criteria for the 44 unIDs considered in this work.

\begin{center}
    \begin{longtable}{|P{3.5cm}|P{2cm}||P{3.5cm}|P{2cm}|}
    \hline 
    \textbf{Source Name} & \textbf{Rejection} & \textbf{Source Name} & \textbf{Rejection}\\
    \hline
    3FGL J0003.4+3100 & F & 3FHL J0301.4$-$5618 & - \\
    3FGL J0244.4+4745 & TS & 3FHL J0302.6+3354 & F \\
    3FGL J0336.1+7500 & - & 3FHL J0343.5$-$6302 & - \\
    3FGL J0538.8$-$0341 & TS & 3FHL J0350.4$-$5153 & X \\
    3FGL J0539.2$-$0536 & F & 3FHL J0359.4$-$0235 & F \\
    3FGL J0600.4$-$1934 & F & 3FHL J0550.9+5657 & TS \\
    3FGL J0905.8$-$2127 & F & 3FHL J0620.9$-$5033 & - \\
    3FGL J0953.7$-$1510 & F & 3FHL J0838.5+4006 & F \\
    3FGL J1106.6$-$1744 & - & 3FHL J0954.2$-$2520 & - \\
    3FGL J1225.9+2953 & - & 3FHL J1403.4+4319 & TS \\
    3FGL J1543.5$-$0244 & - & 3FHL J1421.5$-$1654 & - \\
    3FGL J1936.6$-$4215 & F & 3FHL J1440.2$-$2343 & F \\
    3FGL J1947.4$-$1121 & F & 3FHL J1441.3$-$1934 & - \\
    3FGL J1958.2$-$1413 & TS & 3FHL J1503.3+1651 & - \\
    3FGL J2029.5$-$4232 & - & 3FHL J1650.9+0430 & X \\
    3FGL J2043.8$-$4801 & - & 3FHL J1705.3+5434 & F \\
    3FHL J0041.7$-$1608 & - & 3FHL J1716.1+2308 & - \\
    3FHL J0055.8+4507 & TS & 3FHL J1726.2$-$1710 & TS \\
    3FHL J0110.9+4346 & F & 2FHL J0648.5+4438 & F \\
    3FHL J0115.4$-$2916 & - & 2FHL J0906.8+3530 & F \\
    3FHL J0121.8+3808 & TS & 2FHL J1516.1+3702 & F \\
    3FHL J0233.0+3742 & A & 2FHL J1630.0+7644 & TS \\
    \hline
    \end{longtable}
    \label{tab:rejection_table}
    \begin{quote}
    {\bf Table A} Selection criteria for the 44 unIDs considered in this work. Last column indicates the rejection criteria, if any: TS - the source did not reach the TS=25 detection threshold; F - no annihilation channel was found to fit the SED with $\Delta$TS$\geq$-9; A - associated in 4FGL \cite{4fgl_paper}; X - X-ray counterpart with \textit{Swift}-XRT \cite{Kaur2019}.
    \end{quote}
\end{center} 
\bibliographystyle{JHEP.bst}
\bibliography{References.bib}

\end{document}